\documentclass[fleqn,usenatbib]{mnras}

\usepackage[T1]{fontenc}

\DeclareRobustCommand{\VAN}[3]{#2}
\let\VANthebibliography\thebibliography
\def\thebibliography{\DeclareRobustCommand{\VAN}[3]{##3}\VANthebibliography}
\newcommand{\Bpara}{B$_{\rm{\parallel}}$}
\newcommand{\Bperp}{B$_{\rm{\bot}}$}
\newcommand{\Btot}{{\rm B}_{\rm tot}}

\usepackage{graphicx}	
\usepackage{amsmath,amssymb}
\usepackage{txfonts}
\usepackage{adjustbox}
\usepackage{xcolor}
\usepackage{float}

\title[Magnetic fields on FIRE]{Magnetic Fields on FIRE: Comparing B-fields in the multiphase ISM and CGM of Simulated L$_*$ Galaxies to Observations} 

\author[S. Ponnada et al.]{
\vspace{0.1cm}
\parbox[t]{\textwidth}{Sam B. Ponnada,$^{1}$\thanks{E-mail: sponnada@astro.caltech.edu}
Georgia V. Panopoulou,$^{1}$\thanks{Hubble Fellow}
Iryna S. Butsky,$^{1}$
Philip F. Hopkins,$^{1}$
Sarah R. Loebman,$^{2}$
Cameron Hummels,$^{1}$
Suoqing Ji,$^{3}$
Andrew Wetzel,$^{4}$
Claude-Andr\'e Faucher-Gigu\`ere,$^{5}$
Christopher C. Hayward$^{6}$
}
\\
$^{1}$California Institute of Technology, TAPIR, Mailcode 350-17, Pasadena, CA 91125, USA\\
$^{2}$Department of Physics, University of California, Merced, 5200 N. Lake Road, Merced, CA 95343, USA\\
$^{3}$Astrophysics Division \& Key Laboratory for Research in Galaxies and Cosmology, Shanghai Astronomical Observatory, Chinese Academy of Sciences,\\ Shanghai 200030, China\\
$^{4}$Department of Physics \& Astronomy, University of California, Davis, CA 95616 \\
$^{5}$Department of Physics and Astronomy and CIERA, Northwestern University, 2145 Sheridan Road, Evanston, IL 60208, USA\\
$^{6}$Center for Computational Astrophysics, Flatiron Institute, 162 Fifth Avenue, New York, NY 10010, USA\\
}

\date{Accepted XXX. Received YYY; in original form ZZZ}

\pubyear{2022}

\begin{document}
\label{firstpage}
\pagerange{\pageref{firstpage}--\pageref{lastpage}}
\maketitle

\begin{abstract}
The physics of magnetic fields ($\textbf{B}$) and cosmic rays (CRs) have recently been included in simulations of galaxy formation. However, significant uncertainties remain in how these components affect galaxy evolution. To understand their common observational tracers, we analyze the magnetic fields in a set of high-resolution, magneto-hydrodynamic, cosmological simulations of Milky-Way-like galaxies from the FIRE-2 project. We compare mock observables of magnetic field tracers for simulations with and without CRs to observations of Zeeman splitting and rotation/dispersion measures. We find reasonable agreement between simulations and observations in both the neutral and the ionised interstellar medium (ISM). We find that the simulated galaxies with CRs show weaker ISM $|\textbf{B}|$ fields on average compared to their magnetic-field-only counterparts. This is a manifestation of the effects of CRs in the diffuse, low density inner circum-galactic medium (CGM). We find that equipartition between magnetic and cosmic ray energy densities may be valid at large ($>$ 1 kpc) scales for typical ISM densities of Milky-Way-like galaxies, but not in their halos. Within the ISM, the magnetic fields in our simulated galaxies follow a power-law scaling with gas density. The scaling extends down to neutral hydrogen number densities $<$ 300 cm$^{-3}$, in contrast to observationally-derived models, but consistent with the observational measurements. Finally, we generate synthetic rotation measure (RM) profiles for projections of the simulated galaxies and compare to observational constraints in the CGM. While consistent with upper limits, improved data are needed to detect the predicted CGM RMs at 10-200 kpc and better constrain theoretical predictions. 
\end{abstract}




\section{Introduction}
Magnetic fields are of considerable importance in galaxies, as they are a substantial source of pressure support in the interstellar medium (ISM) and circumgalactic medium (CGM) \citep{Beck2015}. They are capable of significantly influencing the dynamics of both fully-ionised gas and star-forming molecular clouds, thereby modulating star formation rates \citep[for a review, see][]{Krumholz2019}. Magnetic fields also determine the propagation of cosmic rays (CRs) throughout the ISM and into the CGM \citep{Fermi1949,Kulsrud1969,Desiati2014,Shukurov2017}. Despite their well established physical significance, magnetic fields and their connection to galaxy evolution have yet to be fully understood, with progress limited by the ability to accurately characterize magnetic field strengths and topologies. Obtaining accurate measurements of the magnetic field strength and geometry in and around galaxies has implications for many open questions, including their origins and amplification, as well their role in providing non-thermal pressure support and influencing the physical state of the ISM and CGM \citep{Butsky2017,Ji2018, Hopkins2018, Rodrigues2019,Ntormousi2020,Pakmor2020a}. 

Obtaining reliable observational measurements of the field strengths and topologies remains difficult. Most observable tracers of the magnetic field are indirect and rely on certain assumptions: most notably, that of equipartition between CR and magnetic energy densities. 
Assuming equipartition/minimum-energy was first employed to determine field strengths in the jet of M87 by \citet{Burbidge1956}, and has been utilized to determine galactic field strengths \citep{Beck2000,Fletcher2011, Chyzy2011, Beck2015}. 
These estimates utilize the total synchrotron intensity to give information about the magnetic field in the plane of the sky, perpendicular to our line-of-sight (\Bperp), which is used to infer the total magnetic field strength ($\Btot$). 
As \citet{Beck2015} delineates, this method of estimating magnetic field strengths is not without caveats; variation of B along the line-of-sight (LOS) or within the telescope beam \citep{Beck2003}, energy losses of CR electrons \citep{Beck2005}, and invalidity of equipartition on small scales can all lead to overestimating or underestimating the true field strength \citep{Stepanov2014}. 
 
 Other observational measurements of magnetic field strengths are sensitive to the field component parallel to the LOS (\Bpara) convolved with various LOS plasma properties, and uncertain ISM phase structure. These include measurements from Zeeman splitting of spectral lines, which probes the cold, atomic ISM \citep[e.g.,][]{Crutcher2010,Crutcher2011} as well as the use of the ratio of the rotation measure (RM) and dispersion measure (DM) towards background sources like pulsars or fast radio bursts (FRBs) \citep{Han1999, Lan2020, Seta2021}, which probe the highly-ionized phases. Both Zeeman splitting and RM/DM are sensitive to the direction of the magnetic field along the LOS, and measure the magnitude of the regular (ordered and coherent) field component, weighted differently by properties like gas density. It is worth noting that inferring magnetic field strengths from RM/DM relies on the assumption that the thermal electron density and magnetic field strength are uncorrelated \citep{Beck2015,Seta2021}.
 
 A common thread amongst measurements of magnetic fields in galaxies is reliance on several simplifying assumptions which may or may not hold in the multiphase ISM and CGM. This provides motivation for exploring the validity of these assumptions from a theoretical perspective, along with the relative paucity of forward modeled predictions for RM/DM and Zeeman inferred measurements of magnetic fields, especially in cosmological simulations of galaxy formation with amplification from primordial fields. Furthermore, CRs have only recently been included in simulations of galaxy formation, allowing for physics prescriptions which can notably change properties of the ISM and CGM, as well as the potential to forward model synchrotron emissivities \citep{Pakmor2016,Hopkins2020,Werhahn2021,Pfrommer2021}. 

It is only recently that simulations of galaxy formation that included magnetic fields have been capable of following the evolution of galaxies over cosmic time while resolving the ISM  \citep[for example,][]{Marinacci2014,Pakmor2014,Rieder2017,Hopkins2018}. Previously, galaxy-scale simulations that included magnetic field information were often limited to mostly idealized, non-cosmological simulations \citep{Wang2009,Pakmor2013,Su2017,Butsky2017,Steinwandel2019,Steinwandel2020,Steinwandel2022}. Within state-of-the-art simulations, models for feedback and their numerical implementations vary considerably. It has been shown that including essential physics such as gas cooling, star formation, and feedback from stars results in different magnetic field saturation strengths and morphologies \citep{Rieder2017,Su2018}.

Previous work which analyzed magnetic fields includes \citet{Hopkins2020}, who used a set of simulations with magnetic fields and magnetic fields including CRs (which are also analyzed in this study) to demonstrate that magnetic pressure appears to be generally sub-dominant to thermal pressure ($ \beta = P_{\rm thermal}/P_{\rm magnetic} >>$ 1), especially in the CGM, though this study did not closely analyze the magnetic field strengths in the dense ISM, where conditions can be markedly different. They also found that field morphology is tangled on all scales, and found hints of observationally relevant trends with regards to clumping factors, equipartition, and halo gas distributions, but they did not perform a detailed comparison of the magnetic fields with observations. 

Cosmic rays have been found to have little impact on the magnetic field structure and strength in the CGM, however, they have been shown to significantly influence on the dynamics and phase structure of gas in the disk-halo interface, which is the region within 10 kpc vertically from the disk plane \citep{Ji2020,Chan2021}. The resulting impact on the recycling of outflows, i.e. fountain flows, may be of considerable importance to the amplification of magnetic fields \citep{Su2017,AnglesAlcazar2017,Martin-Alvarez2018}. But again, these analyses were not focused on observational comparison.

Despite efforts in understanding the physical implications of magnetic fields and CRs on galactic properties, specifically in the CGM, there have been few forward-modeled observations from idealized and cosmological simulations with explicit treatment of magnetic fields and/or CRs \citep[e.g.,][]{Pakmor2018,VdV2021, Werhahn2021, Pfrommer2021}, and little focus on magnetic fields in the ISM of simulated galaxies \citep[though, see][]{Guszejnov2020, Pakmor2020a, Rappaz2022}.

 In this study, we present analyses of six cosmological `zoom-in' simulations from the Feedback in Realistic Environments Project (FIRE-2\footnote{\url{https://fire.northwestern.edu/}}), described in Section \ref{sec:sims}. We aim to compare synthetic observational tracers of magnetic fields to observed quantities in both the ISM and CGM of L$_*$ galaxies. This is done for two different physical models, one including cosmic rays and one without, and we discuss how the inclusion of CRs impacts the magnetic fields and their observational tracers in Section \ref{sec:analysis}. We also investigate the degree to which the simulated galaxies' magnetic fields and tracers match observations. In Section \ref{sec:discussion}, we compare our results in context of other relevant work, and in Section \ref{sec:conclusions}, we summarize our results and discuss future work on probing galactic magnetic fields through synthetic observations.

\vspace{-20pt}

\begin{table*}
    \centering
    \begin{adjustbox}{max width=\textwidth}
    \begin{tabular}{lcccccccc}
    \hline
    Simulation & M$^{\rm vir}_{\rm halo}$ [M$_\odot$]& M$^{\rm MHD}_{\rm *}$ [M$_\odot$] & M$^{\rm CR}_{\rm *}$ [M$_\odot$] & $<|\textbf{B}|>^{\rm MHD}_{\rm disk}$ [$\rm{\mu}$G] & $<|\textbf{B}|>^{\rm CR}_{\rm disk}$ [$\rm{\mu}$G] & $<|\textbf{B}|>^{\rm CR}_{\rm inner\, CGM}$ [$\rm{\mu}$G] & $<|\textbf{B}|>^{\rm CR}_{\rm outer\, CGM}$ [$\rm{\mu}$G]& Description \\
    \hline
     m12i    & 1.2e12 & 7e10 & 3e10 & 7.93 & 3.99 & 0.025 & 0.012& Late forming MW-mass halo with a massive disc \\
    m12f & 1.6e12 & 8e10 & 4e10 & 5.36 & 4.64 & 0.024 & 0.011& MW-like disk with a LMC-like satellite merger \\
    
    m12m & 1.5e12 & 1e11 & 3e10 & 10.65 & 1.78 & 0.012& 0.008& Earlier forming halo with strong bar at lower redshift \\ 
    \hline
    \end{tabular}
    \end{adjustbox}
    \caption{Simulation properties for the simulations analyzed in this study, from \citet{Hopkins2020}. The columns show the halo's virial mass at $z = 0$ following \citep{Bryan1998}, the galaxy's stellar mass in the MHD+ and CR+ simulations at $z = 0$, the galaxy's average magnetic field strength in the disk at $z = 0$ (defined as the mass-averaged magnetic field strength in all gas cells in the disk), the mass-averaged magnetic field strength in all gas cells in the inner CGM (50 kpc $<$ r $<$ 100 kpc), the mass-averaged magnetic field strength in all gas cells in the outer CGM (100 kpc $<$ r $<$ 240 kpc), and a short description of each galaxy. We show only the CR+ $<|\textbf{B}|>^{\rm CR}_{\rm CGM}$ as at large radii from the galaxy, $<|\textbf{B}|>$ is nearly identical between MHD+ and CR+ simulations (Fig \ref{fig:RM_profiles}).} 
    \label{tab:sims}
\end{table*}

\begin{figure*}
    \includegraphics[width=1.0\textwidth]{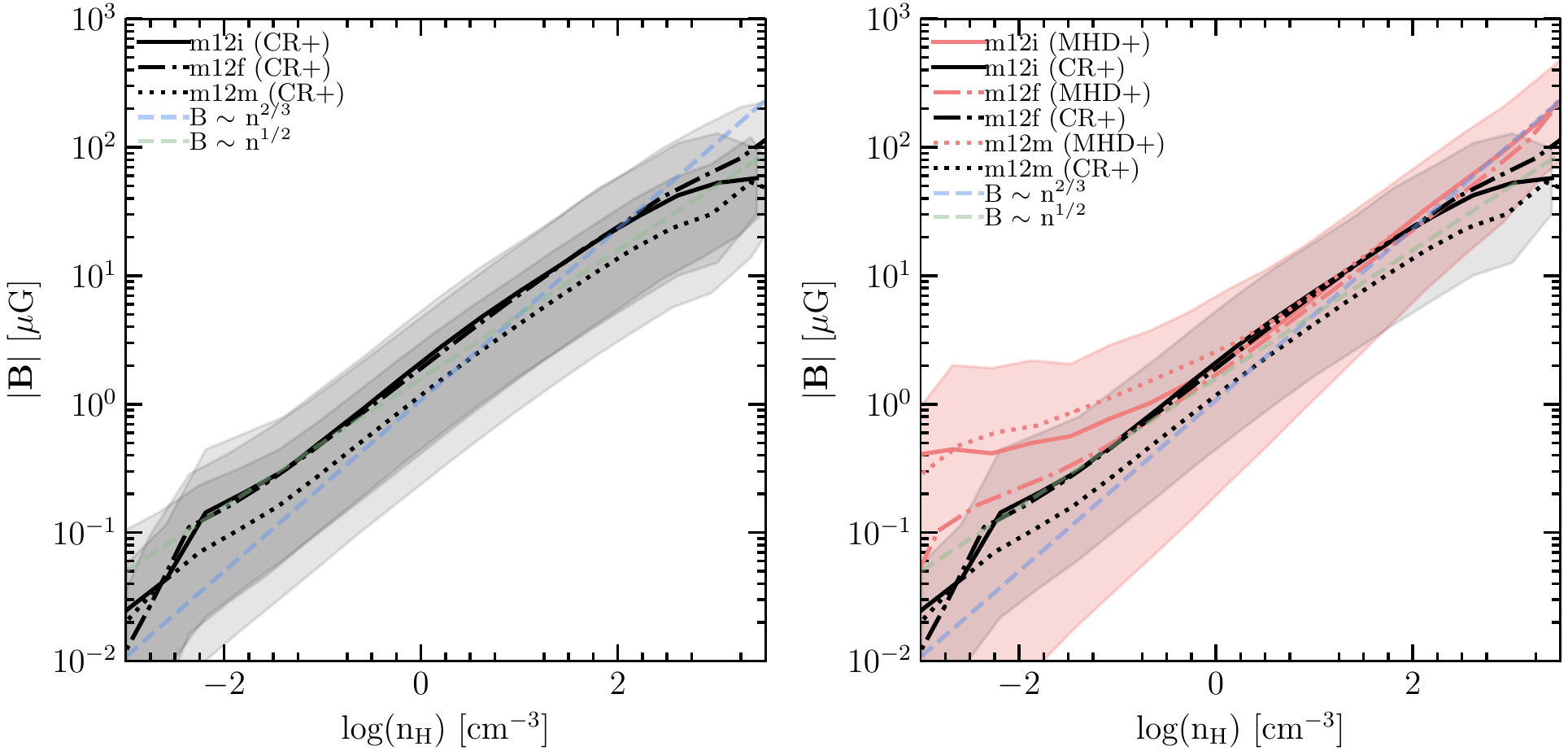}
    \caption{\textit{Mean $|\bf{B}|$ vs. gas number density, $\rm n_H$, for all gas cells in the galactic disk}. \textbf{Left}: Mean $|\bf{B}|$ at each $\rm n_H$ bin for each of the three CR+ simulations (black lines as described in legend). The shaded regions represent the 5-95 percentile range. Scaling relations expected for isotropic flux freezing, with and without spherical geometry of $|\bf{B}|$ $\sim$ n$_H^{2/3}$ and n$_H^{1/2}$ are also shown (blue and green dashed lines, respectively). \textbf{Right}: Mean $|\bf{B}|$ at each $\rm n_H$ bin for m12i (MHD+) in coral, and the same for the CR+ run in black.   Owing to second-order dynamical effects in the "inner CGM," CR+ simulations exhibit suppressed magnetic field strengths at low gas densities ( $\rm n_H \, <$ 0.1 cm$^{-3}$). All of our simulations show roughly the same power-law scaling relation in $|\bf{B}|$ vs. $\rm n_H$, consistent with that of isotropic flux-freezing, with subtle normalization differences due to galaxy-galaxy variation. }
    \label{fig:Bvsdens_wo_obs}
\end{figure*}

\section{Simulations and Methods}\label{sec:sims}
We refer the reader to \citet{Hopkins2020} and references therein for extensive details on the simulations. Here we summarize the most relevant information, and list the fundamental properties of each simulation in Table \ref{tab:sims}. The simulations analyzed here are part of the second iteration of the FIRE project, FIRE-2 \citep[see][]{Hopkins2018}, and so include the physics of gas cooling, explicit treatment of stellar feedback (stellar winds, radiation, and SNe), with the set analyzed in this study including magnetic fields, cosmic rays, and fully anisotropic conduction and viscosity. These simulations are fully cosmological, with adaptive treatment of hydrodynamics and gravity in gas cells, and constant softening parameters for stellar and dark matter particles. The equations of ideal magneto-hydrodynamics (MHD) are solved, and simulations with cosmic rays include an ultra-relativistic fluid ($\gamma$ = 4/3) treatment of CRs with injection from SNe and fully anisotropic streaming, advective and diffusive terms, loss terms, and gas coupling. The injection of CRs from SNe is done by assuming 10 percent of the fiducial SNe energy of 10$^{51}$ erg goes to CRs and is coupled to gas adjacent to the SNe site. Since it is thought that CRs with energies of $\sim$ 1 GeV dominate the CR energy density in L$_*$ galaxies, the CR energy density is evolved for CR energies only around this value \citep{Boulares1990}. In this study, the simulations with cosmic rays assume a constant effective diffusion coefficient $\kappa =$ 3e29 cm$^{2}$ s$^{-1}$, which is observationally motivated \citep{Chan2019}.

We focus only on the most massive galaxy in each of three zoom-in volumes, galaxies roughly akin in mass and size to the Milky-Way, which are named m12i, m12f, and m12m (see Table \ref{tab:sims}). Note that while the runs analyzed in this study are not publicly available, snapshots of the FIRE-2 simulations with the fiducial treatment of feedback physics are publicly available\footnote{\url{http://flathub.flatironinstitute.org/fire}} \citep{Wetzel2022}. Some important systematic trends are notable between the simulations modeling MHD and those modeling MHD including cosmic rays (hereafter denoted MHD+ and CR+, respectively). 
One of these systematic trends is that the CR+ galaxies are $\sim$ 2-3 times lower in stellar mass than their MHD+ counterparts. The reason for the systematically lower stellar masses in the CR+ galaxies is explained in \citet{Hopkins2020, Ji2020, Ji2021} as CR pressure support in the CGM preventing gas for star formation from precipitating onto the disk at redshift $z$ $\lesssim$ 1-2. Correspondingly, the CR+ simulations also tend to have systematically lower ISM gas masses, gas densities, and star-formation rates (SFRs), as well as lower ($\sim$ 20 percent) neutral hydrogen velocity dispersions in the inner disk, $\pm$ 250 pc vertically from the disk mid-plane \citep{Chan2021}.

When we calculate line-of-sight integrated quantities, we project the  galaxy face-on or edge-on using the angular momentum vector of the stars to define the direction perpendicular to the galactic disk, which we use as the z component. Every particle position and vector field in the simulation is transformed accordingly, and this is the coordinate frame in which we define spatial regions below. We integrate gas quantities using the method of \citet{Hopkins2005} along a set of lines of sight (LOSs) which uniformly sample the desired area. The sampling is done by dividing the area of interest into a two-dimensional, 700 $\times$ 700 image, with little to no difference in the results when increasing the image resolution. The required magnetic fields and gas quantities such as temperature and ionization state are calculated self-consistently in-code \citep{Hopkins2015,Hopkins2018,Hopkins2020}. 
 
For ISM quantities, we restrict LOS integrations to gas within a cylindrical region of $14\,\rm{kpc}$ to account for the extent of our galaxies' gas disks \citep{Bellardini2021} and height $\rm |z|<1$\,kpc, though none of our results are especially sensitive to the exact threshold. If we restrict to "Solar Circle" radii, we only include cells with galactocentric radii, $R$, that satisfy: $7\,{\rm kpc} < R < 9\,{\rm kpc}$. Furthermore, since the Milky-Way pulsar observations we compare to (\ref{sec:pulsars}) sample typical distances to sources between $\sim 0.1-4$\,kpc, we divide the ISM integration into "slabs" uniformly sampling varying depths in log(distance) over this range; we agglomerate a collection of sightlines through the disk with slabs of thickness in z of 0.1, 0.2, 0.5, 1, 2, and 4 kpc. This is equivalent to placing sources and observers randomly in order to sample a similar distance distribution to the observations. For CGM quantities, we integrate sightlines through a sphere of physical radius $ r =$ 245 kpc, corresponding to a projected radius of 200 kpc centered on the galaxy. We define the physical radius to be $r = \sqrt{x^2 + y^2 +z^2}$, while the projected galactocentric radius is $R = \sqrt{x^2 + y^2}$ for a face-on projection and $R = \sqrt{x^2 + z^2}$ for an edge-on projection (where x, y, z are all centered on the galaxy center).

\section{Results}\label{sec:analysis}

\subsection{Magnetic Fields vs. Gas Density in the ISM}\label{sec:Bvsn}

We first investigate the relation between the  magnetic field, {\bf{B}}, and gas density, $\rm n_H$, in the simulations, within the ISM. Throughout this work, we often refer to the magnetic field in the simulation as the "true" magnetic field to distinguish from observationally-derived magnetic field measures. We select all gas cells in the disks, i.e. cells with $\rm z$ coordinate $\rm |z| < 1$ kpc and galactocentric radius $R = \sqrt{x^2 + y ^2} < 14$ kpc, where $\rm x, y$ refer to the coordinates of the cell in the simulation volume. 

The relation of {$|\bf{B}|$} with $\rm n_H$ for the different MHD+ and CR+ simulations is shown in Figure \ref{fig:Bvsdens_wo_obs}. In our simulations, $|\textbf{B}|$ $\sim$ n${\rm _H}^{\alpha}$ for typical ISM gas densities ($\rm n_H \gtrsim 1 cm^{-3}$), where $\alpha$ is a power-law fit to the mean and ranges around $\sim$ 0.5-0.6. In a simplistic case where the exponent $\alpha$ is set by the collapse of gas in the absence of magnetic flux diffusion (flux freezing), $\alpha$ is related to the geometry of the collapse. The value of $\alpha$ can range from 0, if gas collapses along magnetic field lines, to 1, if the collapse is perpendicular to field lines, while the case of isotropic collapse gives an exponent of $\alpha = 2/3$ \citep{Crutcher2010,Crutcher2011,Tritsis2015}. Our simulations exhibit a scatter in the slope of the relation consistent with the resulting trend from isotropic flux freezing, $|\textbf{B}|$ $\sim \rm n_H^{2/3}$ (or $|\textbf{B}|$ $\sim \rm n_H^{1/2}$ in the case where the assumption of spherical cloud geometry is invalid. $|\textbf{B}|$ $\sim \rm n_H^{1/2}$ is preferred where the cloud geometry is more slab-like or filamentary, with field lines perpendicular to the slab, or at an angle relative to the primary axis of the filament, which collapses radially \citep[see][for a detailed discussion]{Tritsis2015}.

At lower densities, $\rm n_H \ll 1 cm^{-3}$, we find different behaviors for the MHD+ and CR+ simulations. While the mean values of $|\textbf{B}|$ in the MHD+ and CR+ relations mostly agree at typical ISM densities (the mean ISM number density for these simulated galaxies is $\sim$ a few cm$^{-3}$, with a standard deviation of $\sim$ 100 cm$^{-3}$), there is a substantial offset (of a factor of 2-10) in the very diffuse gas, at number densities less than 0.01 cm$^{-3}$. For m12i and m12m, shown in the right panel of Figure \ref{fig:Bvsdens_wo_obs}, the CR+ simulations exhibit suppressed mean $|\textbf{B}|$ values compared to the MHD+ simulations. For m12f, however, there is no significant offset. When comparing among the CR+ runs, we find only modest galaxy-to-galaxy differences in their {$|\bf{B}|$} -- $\rm n_H$ relations. While m12i (CR+) and m12f (CR+) exhibit very similar trends in $|\textbf{B}|$ vs. $\rm n_H$, we find that m12m (CR+) has a mean $|\textbf{B}|$ systematically lower at a given gas density at $\rm n_H > 1 \, cm^{-3}$ by a factor\footnote{While this offset at higher gas densities is noteworthy, we are cautious about how much to interpret this, given that it is much smaller than the instrinsic scatter in $|\bf{B}|$ at a given number density.} of $\sim$ 2. The origin of these galaxy-to-galaxy variations may arise in the specific merger histories of the galaxies, as m12m has a considerably different merger history than m12i and m12f, with m12m (CR+) also exhibiting the largest difference in stellar mass from its MHD+ counterpart, or due to its different gas and stellar mass profiles and morphology \citep[see][]{Hopkins2020}, but we are unable to identify any single variable that explains the more systematic offset in this galaxy. Investigating this is beyond the scope of the present work, however, for our purposes, these observed variations can serve as rough guide for "systematic" uncertainties present in our derived $|\textbf{B}|$ vs. $\rm n_H$ relations. 

From our comparison between the CR+ and MHD+ simulations, it is clear that at typical ISM densities in the warm ionised medium (WIM), warm neutral medium (WNM), cold neutral medium (CNM), and molecular phase, cosmic rays do not alter typical magnetic field strengths {\em at a given gas density}, though m12m CR+ exhibits systematically lower field strengths, this owing more to galaxy-galaxy variation than to cosmic ray effects. While CRs and magnetic fields are known to influence each other via plasma instabilities \citep[e.g.][]{Bell2004}, these effects are well below our resolution scale ($\sim$ 66 pc at $\rm n_H$ of 1 cm$^{-3}$, see \citet{Hopkins2018}), and have very little effect on galaxy-scale magnetic fields. Thus, while they may influence the "average" magnetic field strength within a galaxy, it would be indirectly, through changing the overall mass budget at different gas densities or in different phases, moving along the same {$|\bf{B}|$} -- $\rm n_H$ relation. 

We have demonstrated that in some cases, CRs do appear to indirectly lower magnetic fields in the lowest-density gas. We have confirmed that this offset is not a result of the overall mass offset between CR+ and MHD+ simulations, by comparing the "m11" simulations (order of magnitude lower-mass halos) run with MHD+, which do not exhibit such an offset in field strength at a given gas density. Moreover, inspection shows that the offset seen in Fig. \ref{fig:Bvsdens_wo_obs} (right) is not coming from supernova bubbles (the HIM), nor from any particular position within the disk midplane (measuring these trends just at the solar circle versus averaged over the whole disk shows the same effect). And we show below that in the outer CGM (far from the galaxy), magnetic field strengths are roughly the same in the MHD+ and CR+ simulations. The systematic difference appears to manifest primarily in the "inner CGM" or "disk-halo interface" -- tenuous gas between the midplane and $\sim$\,a few kpc above the disk. In Section \ref{sec:discussion}, we discuss how this is not a direct effect of including CR physics, but rather a second-order effect of the dynamical influence of CRs in this region.

\begin{figure*}
    \centering
    \includegraphics[width=1.0\textwidth]{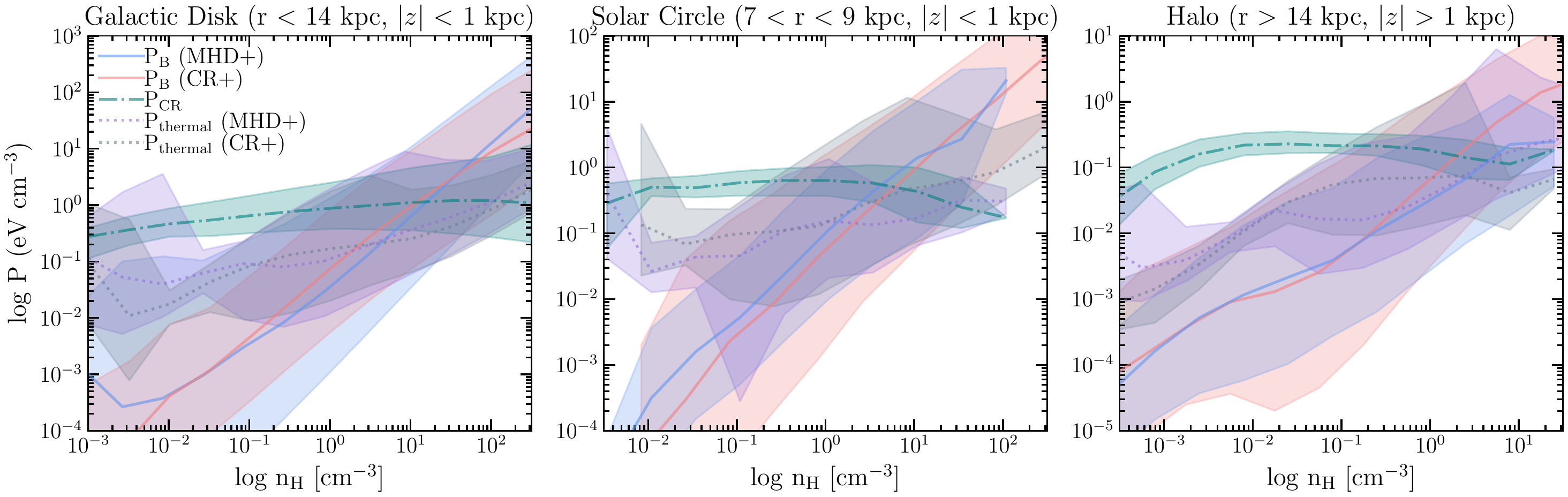}
    \caption{\textit{Pressure budget as a function of gas density, for various regions in the simulations of m12i.} Here, we show magnetic pressure, $\rm P_B$, with blue and coral solid lines (for MHD+ and CR+, respectively), thermal pressure, $\rm P_{thermal}$, with purple and gray dotted lines (MHD+ and CR+, respectively) and CR pressure, $\rm P_{CR}$, with teal dash-dotted line for the CR+ run. The lines represent the means for each density bin and shaded regions show 5-95 percentile intervals (approximate 2$\sigma$). \textbf{Left:} galactic disk, \textbf{center}: approximate solar circle, \textbf{right}: halo/CGM. In the galactic disk, cosmic rays are in pressure equilibrium with magnetic energies at typical ISM densities (n$_{\rm H}$ $\sim$ 1-10 cm$^{-3}$), indicating that equipartition assumptions may hold. In the halo, however, magnetic fields are subdominant to cosmic rays in the diffuse phase which fills the CGM volume, and not in equilibrium with thermal pressure.}
    \label{fig:m12i_equi}
\end{figure*}

\subsection{Pressure budget of the ISM and CGM}

In this section, we examine the density dependence of energy densities in the simulations, allowing us to see in which regions of parameter space the magnetic pressure can dominate. Notably, in our CR+ simulations, the cosmic ray energy density ($e_{\rm CR}$) at 1 GeV is evolved self-consistently, allowing us to examine whether commonly adopted assumptions about equipartition between magnetic and CR energy densities hold in these simulations.

In Figure \ref{fig:m12i_equi}, we present the magnetic, thermal, and CR pressures (defined as P$_{\rm B}\equiv \rm |B|^{2}/8\pi$, P$_{\rm thermal} \equiv \rm n_H k_{B} T$, P$_{\rm CR} \equiv (\gamma_{CR}-1)\, e_{\rm CR})$, respectively) as a function of n$_{\rm{H}}$ for different regions of m12i. Note that P$_{\rm B}$ is also equivalent to u$_{\rm B}$, the magnetic energy density. At densities of around 1-10 cm$^{-3}$, the magnetic energy density is in approximate equipartition with that of CRs, especially in the "Solar circle" (Fig. \ref{fig:m12i_equi}, middle). This is in agreement with observational constraints in the Solar neighborhood \citep{Strong2000,Beck2001}. Equipartition between these various pressures does not hold generally; in our simulations, we see that the CR pressure dominates in the low density ($\rm n_H$ $<$ 1 cm$^{-3}$) gas. This can be understood by the fact that CRs are able to diffuse across field lines, and are thus more weakly functions of the gas density compared to magnetic and thermal pressures. Furthermore, the exact $\rm n_H$ at which the magnetic and CR pressures are equal varies with distance from the galactic center, comparing the "Solar circle" and halo regions to the whole disk. Notably, when considering all cells in the galactic disk of m12i (CR+), we find that a mean log$_{10}$($\frac{P_{B}}{P_{CR}}$) of $\sim$ -1, with a standard deviation $\sim$ 1.

In the halo, however, magnetic fields do not reach equipartition strengths except at higher densities than the majority of gas in the CGM. Thus, in CR-pressure dominated haloes like those of the FIRE galaxies (see also \citet{Ji2020}) one would overestimate the magnetic field strength when using the common assumption of equipartition applied to observations. In our simulations, the CR energy density is much larger than the magnetic energy density in the halo. For a constant diffusivity (as these simulations assume), the CR energy density decreases as $\sim$ r$^{-1}$ far from the galaxy, while for an isothermal halo with flux freezing, $|{\bf B}|$ $\sim$ r$^{-4/3}$, falls more rapidly \citep{Ji2020}. But we caution that this is of course sensitive to the assumptions of how CRs propagate - the diffusivity may not be constant in nature, and many models that fit the MW solar system constraints equally well can produce a more-rapidly declining P$_{\rm CR}$ with radius (see \citet{Hopkins2021b}).

Furthermore, in the halo, it appears that magnetic fields are in near-equipartition with the thermal pressure at densities of 1-10 cm$^{-3}$, but for diffuse halo gas which fills most of the volume, the magnetic fields are subdominant to thermal pressure. This is consistent with previous studies which found plasma $\beta$ $>>$ 1 in halo gas \citep{Su2018,Butsky2020,Ji2020}, but is in contrast to results found by \citet{Pakmor2020a}, which may have more to do with the differing feedback models and numerical implementations, as we discuss further in Section \ref{sec:discussion}. The relation between thermal and magnetic energy densities in our simulations is in contrast to models which predict fields at equipartition pressure with thermal pressure terms, thought to aid in accretion of cool CGM gas \citep{Pakmor2016,Butsky2018, Prochaska2019a}. At high gas densities ($> \sim$ 50 cm$^{-3}$), magnetic pressures dominate over thermal pressures, but perhaps not over the turbulent/dispersion pressure, which was explored in the ISM of FIRE-2 simulations by \citet{Gurvich2020}.

\begin{figure}
    \centering
    \includegraphics[width=0.48\textwidth]{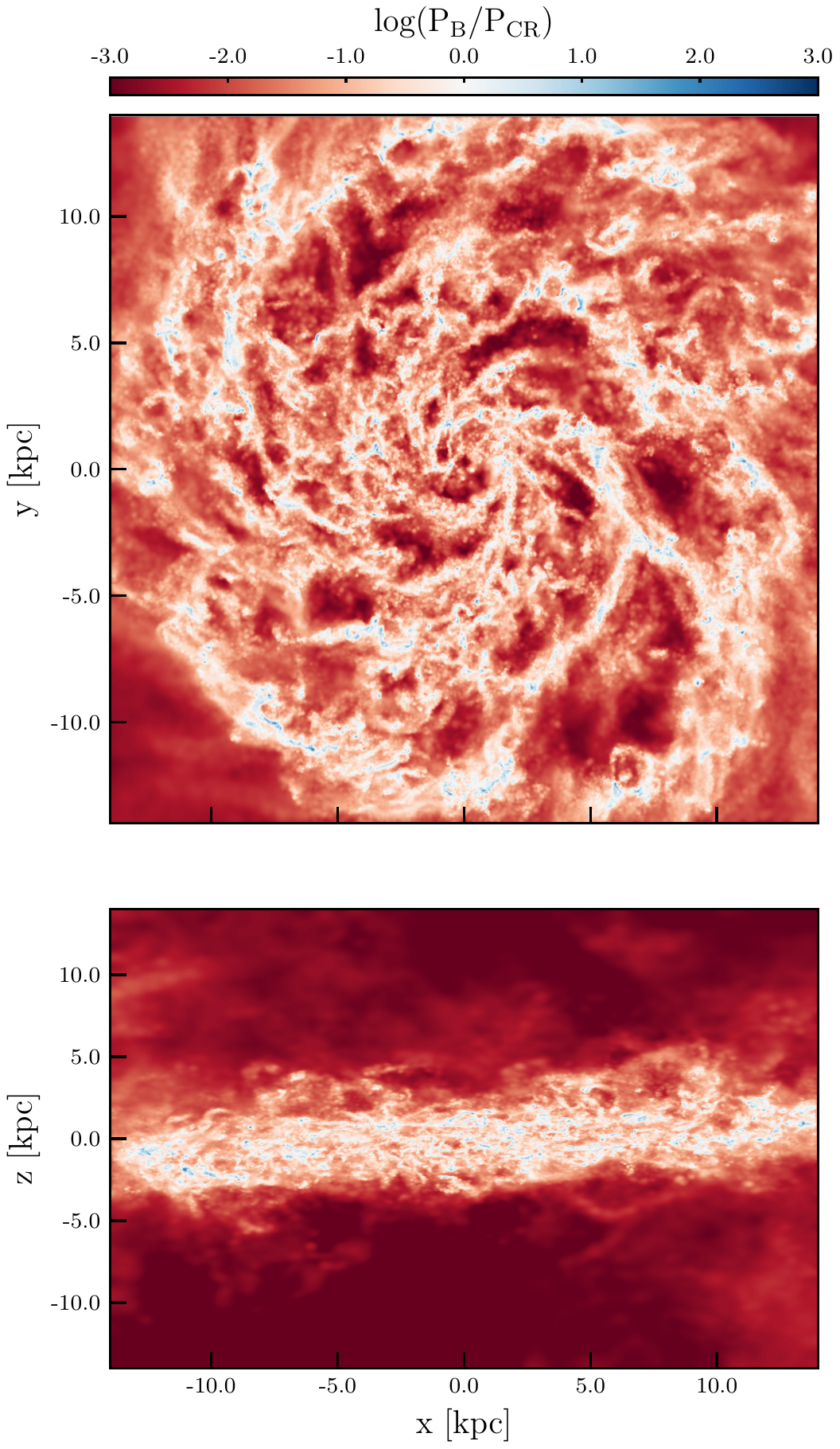}
    \caption{\textit{Logarithm of the ratio between magnetic and cosmic ray pressures in m12i (CR+)} ($\frac{\int P_{\rm B} d\ell}{\int P_{\rm CR} d\ell}$), for a face-on projection in the top panel and edge-on projection in the bottom panel. We see that equipartition (white regions) holds on large ($>$ 1 kpc) scales in the galactic disk, cospatially with spiral structure, however with large deviations in inter-arm regions (dark red pockets). On small scales, equipartition does not hold in regions of high magnetic energy density within dense, molecular gas (small, blue regions).}
    \label{fig:equi_viz}
\end{figure}

Equipartition between cosmic rays and magnetic fields is an assumption whose validity may be scale-dependent due to the propagation of cosmic-rays in the ISM from their injection sites at SNe \citep{Beck2015,Seta2019}. In Figure \ref{fig:equi_viz}, we examine the scale-dependence of equipartition between cosmic rays and magnetic fields in the ISM and disk-halo interface (the very inner CGM) of m12i (CR+) by visualizing the ratio of mass-weighted projections of the magnetic and cosmic ray pressures ($\frac{\int P_{\rm B} d\ell}{\int P_{\rm CR} d\ell}$). These projections are computed using the routine for line-of-sight integrated quantities described in Section \ref{sec:sims}, where for the face-on projection we utilize the same convention described therein, and for the edge-on visualization we integrate the line of sight along $\pm$ 14 kpc from the galactic center.

We find that in the ISM, equipartition is valid on large ($>$ 1 kpc) scales, consistent with observational constraints \citep{Stepanov2014} and recent theoretical work using idealized simulations of magnetized disks \citep{Rappaz2022}. However, this equipartition is not universal, and regions where equipartition holds are primarily cospatial with the density distribution in spiral structures, with large-scale cavities where the ratio of pressures is as low as 10$^{-3}$. On small scales, pressure ratios as high as 10$^{3}$ are apparent, coincident with dense molecular complexes along the line of sight.

\subsection{Magnetic fields in the cold, neutral/molecular ISM: comparison to Zeeman Splitting}
\label{sec:Zeeman}

In this Section, we compare the magnetic field amplitude in our simulations to observations of cold neutral and molecular gas in the Milky Way.

\subsubsection{Line of Sight Magnetic Field Strength vs. \ion{H}{I} Column Density}

The most comprehensive compilation of observations of magnetic field strengths in the cold phases of the ISM has been presented by \citet{Crutcher2010}. These estimates rely on observations of the Zeeman effect, which can be used to probe the line-of-sight component of the magnetic field towards diffuse HI clouds and giant molecular clouds (GMCs).

To directly compare with these observations, we integrate the line-of-sight (parallel) component of the magnetic field in the simulations, B$_
{\rm \|}$, weighted by the cold, neutral hydrogen mass, to obtain a "Zeeman-inferred" magnetic field. Specifically, we construct sightlines through a face-on projection of the disk, and use the routine described in Section \ref{sec:sims} of \citet{Hopkins2005} to calculate the Zeeman-inferred magnetic field as:
\begin{equation}\label{eq:LOS}
    \rm{B_{\rm \| \,, Zeeman-inferred}} = \frac{\int_L {\rm n_{ HI,cold}} \, \rm{B_\|} \, d\ell}{\int_L {\rm n_{ HI,cold}} \, d\ell},
\end{equation}
where the integration is performed along, $\ell$, the path length through the disk. 
The cold fraction observed in Zeeman absorption is estimated as ${\rm n_{ HI,cold}}$ $\sim$ $\rm n_{H} \, e^{\rm -T/50 \,K}$ following \citeauthor{Crutcher2010}, but our results are insensitive to this threshold temperature for any values between $\rm T = 50 - 500$ K as for these cutoff temperatures, the sightlines are still effectively weighted by the dense gas along the line of sight.

\begin{figure}
    \centering
    \includegraphics[width=0.48\textwidth]{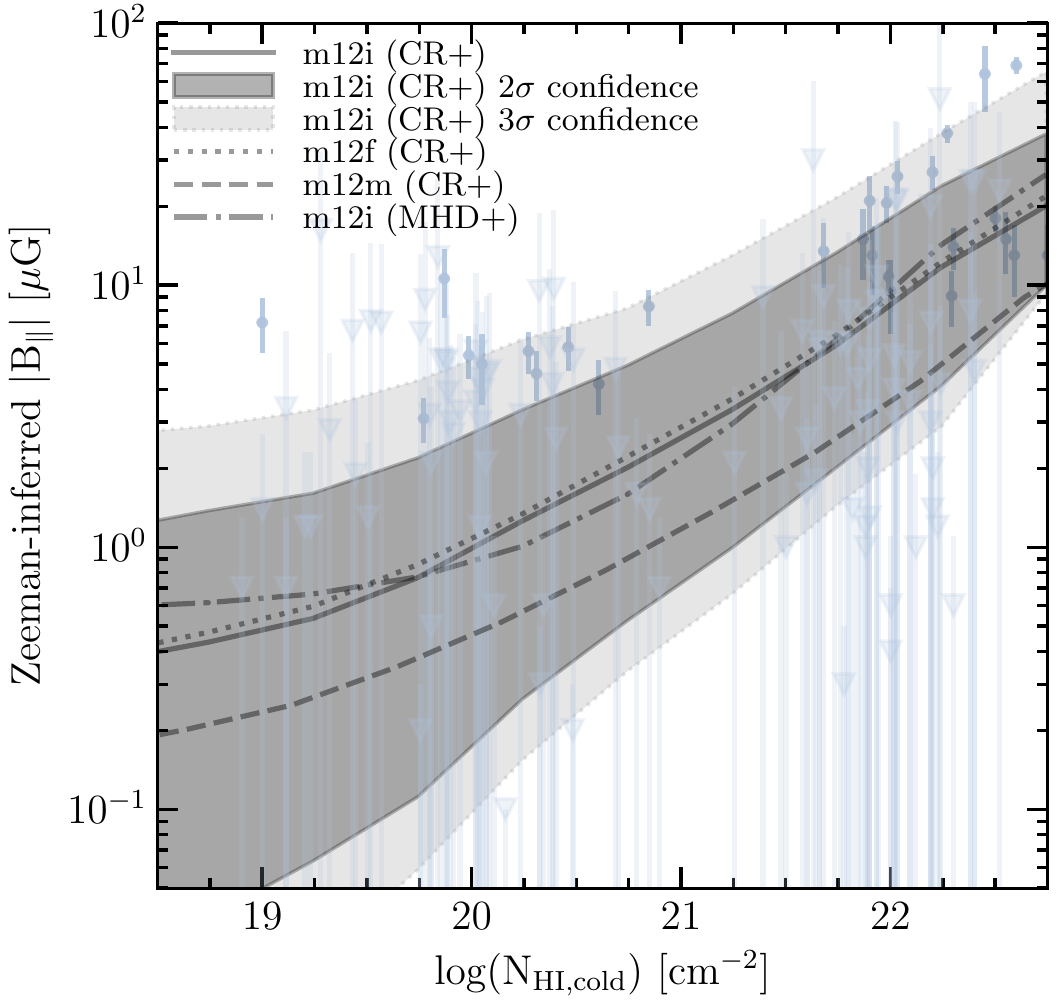}
    \caption{\textit{$\rm{B_{\rm \| \,, Zeeman-inferred}}$ (Equation \ref{eq:LOS}) vs. cold neutral-hydrogen column density} for face-on projections of m12i, m12f, and m12m (CR+). The black solid line shows the median value at each column density bin for sightlines passing through the galactic disk for m12i, with the same for m12f (CR+) and m12m (CR+) shown with dotted and dashed lines. The results are unchanged if restricted to the solar circle. Shaded regions show 5-95 percentile and 1-99 percentile (approximately 2$\sigma$ and 3$\sigma$ confidence regions) for m12i (CR+). Light blue points show observational measurements from \citet{Crutcher2011} and associated 1$\sigma$ error-bars. Non-detections are shown as inverted triangles with +3$\sigma$ upper error-bars. Our CR+ simulations show good agreement with the observational data, however; they are slightly more discrepant at the low column density end due to sightlines probing more diffuse atomic gas, which exhibit lower field strengths compared to the MHD+ simulations.}
    \label{fig:BvsNH}
\end{figure}

Figure \ref{fig:BvsNH} shows the relation between B$_{\rm \| \,, Zeeman-inferred}$ and column density (N$_{\rm{HI}}$) for the CR+ simulations as well as the observational data.
Here, the N$_{\rm{HI}}$ of interest is the same cold, neutral-hydrogen column density, $\int {\rm n}_{\rm HI, cold} \, d\ell$ . The variables required to compute these line-of-sight values are self-consistently calculated in the simulation and require no further modeling. We have compared a set of sightlines computed using the same method uniformly sampling the entire galactic disc, versus those only sampling the solar circle (7-9 kpc), and find that the results at a given N$_{\rm{HI}}$ are nearly identical.

We find that there is reasonable agreement between the Zeeman-inferred magnetic field and the observations across the range of column densities; i.e., the majority of the Zeeman observations lie within the scatter of the mock Zeeman measurements for both the CR+ (Fig. \ref{fig:BvsNH}) and MHD+ (not shown) simulations. However, there is qualitatively poorer agreement at low N$_{\rm{HI}}$, especially in m12m (CR+), where the mock Zeeman measurements lie below the majority of the observations. We attribute this to an overall decrease in the total magnetic field strength relative to the MHD+ simulations, which was shown in Figure \ref{fig:Bvsdens_wo_obs}, due to sightlines probing more diffuse atomic gas \citep[see][who show that the CR+ simulations do feature significantly more cold atomic gas at low densities ${\rm n_H}$ $\sim$ 0.1 - 1 cm$^{-3}$, due to CR pressure support leading to lower thermal pressures at a given gas temperature]{Ji2020}.

It is also important to note that observations do not probe these diffuse sightlines very robustly, and the constraints in the low N$_{\rm{HI}}$ regime are mostly upper-limits.

There are some caveats to these mock observations: these are not full radiative transfer calculations of the Zeeman splitting of spectral lines in individual clouds, nor is any of our simulations a perfect analog to the Milky Way. However, the comparisons shown here demonstrate that the simulated galaxies reproduce the observed Milky Way magnetic field strength - column density relation at the order-of-magnitude level.

\subsubsection{Zeeman-Inferred Field Strengths and Gas Densities}

\begin{figure}
    \centering
    \includegraphics[width=0.48\textwidth]{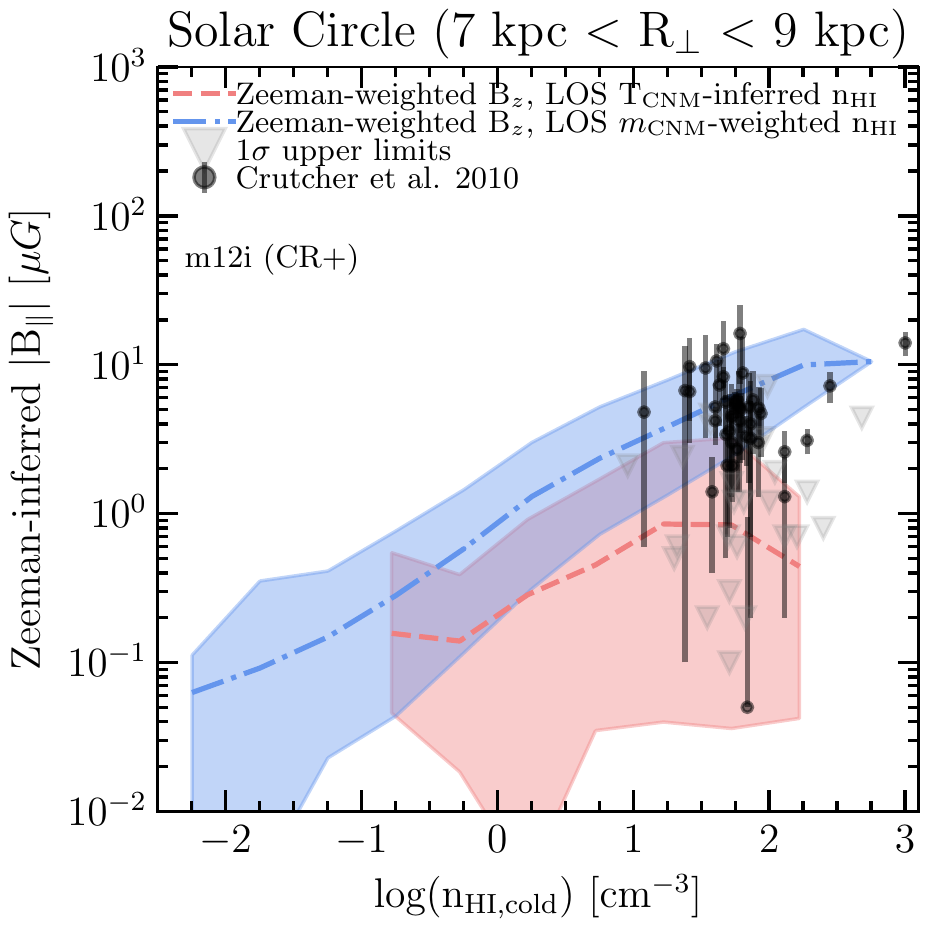}

    \caption{\textit{Zeeman-inferred $|\rm{B_{\rm \|}}|$ vs. cold, neutral hydrogen number density} for m12i (CR+). Each line shows the mean Zeeman-inferred $|B_{\rm \|}|$ at each number density bin. The coral line shows $|B_{\rm \|}|$ vs. n$_{\rm HI,\,T_{\rm CNM}}$ (Equation \ref{eq:nhi_tcnm}), which is the number density inferred from the LOS cold, neutral hydrogen weighted-temperature assuming thermal pressure equilibrium. The blue line also shows the Zeeman-inferred $|\rm{B_{\rm \|}}|$, but plotted against n$_{\rm HI,M_{\rm CNM}}$ (Equation \ref{eq:nhi_cnm}), which is the LOS cold mass-weighted density. Black points show observations from \citep{Crutcher2010} and associated 1$\sigma$ errorbars, and non-detections are shown with inverted grey triangles. Shaded regions show the 5-95 percentiles (approximately 2$\sigma$). The results are nearly identical for the MHD+ run, and for m12f and m12m (not shown). The relation between $|\rm{B_{\rm \|}}|$ and n$_{\rm HI,M_{\rm CNM}}$ is broadly consistent with the purely theoretical values of $|\bf B|$ vs. n. However, when using n$_{\rm HI,\,T_{\rm CNM}}$, the relation is flattened due to thermal pressure equilibrium being a poor approximation in the cold ISM, thus not faithfully tracing the "true" density.}
    \label{fig:Bvsdens}
\end{figure}

In this section, we examine the relation between the Zeeman-inferred $|\rm{B_{\rm \|}}|$ and the three-dimensional gas density which is often discussed in the star formation literature \citep[for a review see][]{Crutcher2011}. Particularly of interest are the results of \citet{Crutcher2010} examined in the previous section, now in a different observational plane which depends on methodology for determining three-dimensional cloud densities. In Figure \ref{fig:Bvsdens}, we study the relation between Zeeman-inferred magnetic field strength and three-dimensional gas density at the Solar circle in our simulated galaxies.

It is not usually possible to determine the mean density $n_{\rm HI,\,cold}$ of the individual Zeeman absorbers from observations, and since it is well-known that the absorbers are small structures along the line-of-sight, using a quantity like $\int_L {\rm n}_{\rm HI,\,cold}\,d\ell / \int_L d\ell$ would severely under-estimate their typical densities. So instead, we compare two approaches. First, a root-mean-squared (rms) or cold-gas mass weighted density, similar to what one might (ideally) estimate via cold gas line emission, which appropriately weights for clumping along the line of sight:

\begin{equation}\label{eq:nhi_cnm}
    {\rm n}_{\rm HI,M_{\rm CNM}} = \frac{\int_L {\rm n}_{\rm HI,cold}^{2} d\ell}{\int_L {\rm n}_{\rm HI,cold} d\ell}
\end{equation}

Second, we approximate the method from \citet{Crutcher2010} and \citet{Crutcher2011}, which assumes that the \ion{H}{I} spin temperature $\rm T_{\rm spin}$ gives the kinetic temperature $\rm T = T_{\rm spin}$ of the gas, which itself is in thermal pressure equilibrium with a universal constant pressure, so that everywhere ${\rm n}_{\rm HI,\,cold} \, \cdot  \rm T = 3000\,{\rm K\,cm^{-3}}$. This gives

\begin{equation}\label{eq:nhi_tcnm}
    {\rm n}_{\rm HI,\,T_{\rm CNM}} = 3000\,{\rm K\,cm^{-3}}\,\frac{\int_L {\rm n}_{\rm HI,\,cold}\,d\ell}{\int_L {\rm n}_{\rm HI,\,cold}\,{\rm T}\,d\ell }
\end{equation}

Figure \ref{fig:Bvsdens} shows the B$_{\rm \| \,, Zeeman-inferred}$ vs density for the two different density estimators. We see that B$_{\rm \| \,, Zeeman-inferred}$ versus $\rm n_{\rm HI,\,M_{\rm CNM}}$ gives at least a broadly similar trend to the "true" theoretical $|{\bf B}|$ versus $\rm n_H$ in Figure \ref{fig:Bvsdens_wo_obs}, with a differing normalization by a factor of $\sim 2$ owing to the geometric effect of measuring solely the line-of-sight component of the magnetic field. While B$_{\rm \| \,, Zeeman-inferred}$ is consistent with the observed data, the relation between B$_{\rm \| \,, Zeeman-inferred}$ and $\rm n_{\rm HI,\,T_{\rm CNM}}$ is flattened significantly, and we find that this effect remains even if we isolate the sightlines with dense complexes ($\rm n_H \, >$ 300 cm$^{-3}$) along the path length, or draw sightlines through those dense complexes themselves, rather than sampling sightlines through the entire disk at the Solar circle. This is because there is very little correlation, over the dynamic range here, between $\rm n_{\rm HI,\,T_{\rm CNM}}$ and either $\rm n_{\rm HI,\,M_{\rm CNM}}$ or the true gas density $\rm n_H$ from which the Zeeman absorption originates. The poor correlation owes to the fact that: (1) \textit{thermal pressure equilibrium is not a particularly good approximation in the cold ISM}, where other forms of pressure (including magnetic as we show in Figure \ref{fig:m12i_equi}, turbulent, and cosmic ray) all dominate over thermal, and the conditions are highly dynamic \citep{Grudic2021}, and (2) even if thermal pressure equilibrium were reasonable, \textit{there is not a single thermal pressure across all clouds in all locations in the Milky Way} (e.g. towards the galactic center, it is well-known that cloud pressures are much higher). As a result, using $\rm n_{\rm HI,\,T_{\rm CNM}}$, which not only assumes constant thermal pressure, but a \textit{single value} of thermal pressure across all sightlines, introduces considerable scatter in the x-axis of Figure \ref{fig:Bvsdens}, essentially "smearing out" the correlation. Recall that our values of $\rm |B_{\|}|$ are the same for all sightlines, so the effect on the normalization depends on which densities are most heavily sampled. If we sampled denser true sightlines more numerously, as in the \citet{Crutcher2010} analysis, then the normalization of this curve would be the same as their favored value.

\subsection{Magnetic fields in the ionised ISM and CGM: comparison to rotation and dispersion measures}
\label{sec:pulsars}

\subsubsection{Comparison to Pulsar RMs and DMs in the Galaxy}
In this Section, we compare the simulations with observational probes of the magnetic field in the ionized medium, primarily through use of Faraday rotation and dispersion measure. Faraday rotation occurs due to the the interaction of light with a magneto-ionic plasma, causing the plane of polarization to rotate as a function of the electron density and magnetic field strength. The average magnetic field along the line of sight ($|\langle B_{\rm \|} \rangle|$) can be determined through the rotation measure if an independent constraint is found on the electron density, $\rm n_e$, given by the dispersion of radio pulses. The dispersion measure is dependent on the thermal electron column density and we compute the DMs along the mock line-of-sight as DM = $\int \rm n_{\rm e} d\ell$. The RMs are computed with the standard definition \citep{Beck2015}:
\begin{equation}\label{eq:RM}
    \rm{RM} = \frac{e^3}{2\rm{\pi m_e^2 c^3}} \int_{L} n_e \textbf{B} \cdot \textbf{dl} = 0.808 \int_{L} n_{\rm e} (cm^{-3}) \textbf{B} (\mu G) \cdot \textbf{dl} (pc),
\end{equation}
where $\rm e$ is the electron charge, $\rm m_e$ is the electron mass, $\rm c$ is the speed of light, and the integration is performed along the path-length $\ell$. 

To compare the synthetic measurements as expected from our MW-analogs to MW pulsar observations, we use the same line-of-sight computation routine described in Section \ref{sec:sims} of dividing the galactic disk into "slabs" to determine RMs and DMs. For Zeeman splitting, it makes no difference if we integrate arbitrarily large sightlines "outside" of the disk because these have a very small probability of intercepting cold atomic \ion{H}{I}. However, for RM/DM measurements, it is important that we sample sightlines of appropriate depths through the disk (similar to those for the actual observed Galactic pulsars), and not simply integrate all sightlines to $\pm\infty$, because then the predicted DM would be completely dominated by the cumulative contribution from halo and IGM ionized gas.

\begin{figure}
    \centering
    \includegraphics[width=0.48
    \textwidth]{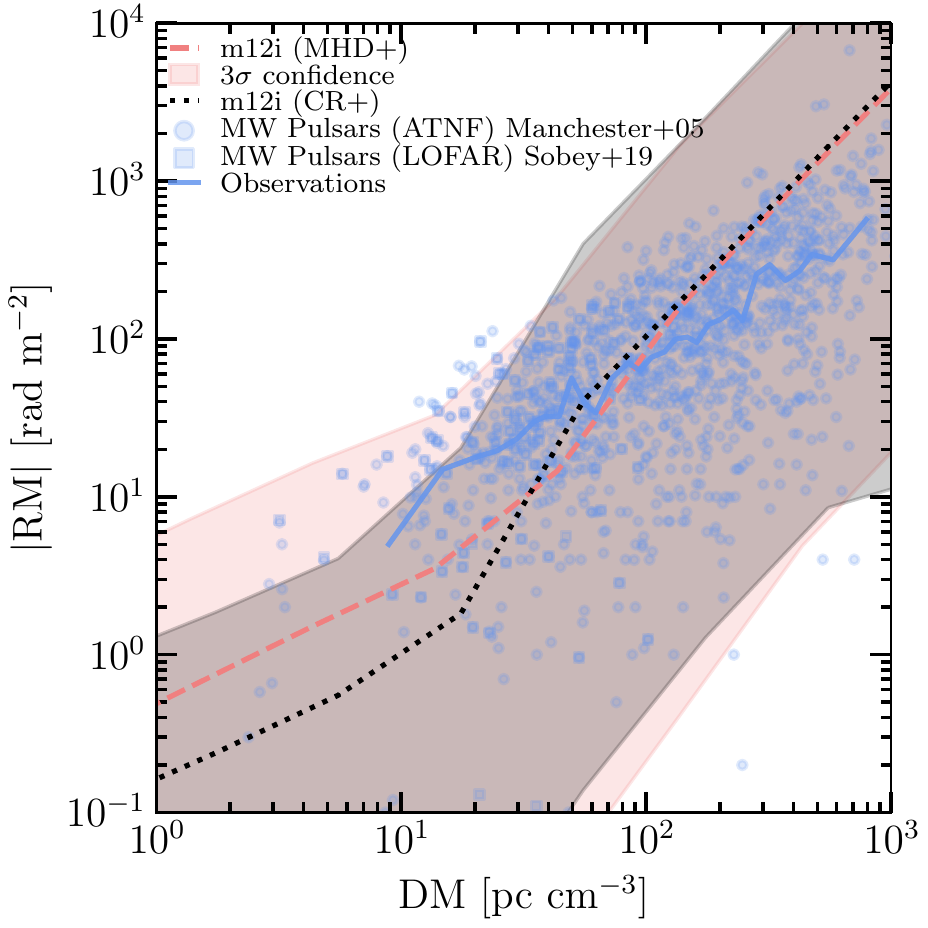}
    \caption{ \textit{$|\rm{RM}|$ vs. DM for sightlines through the disks} of m12i MHD+ and CR+ (coral and black respectively). Lines indicate the mean $|RM|$ in each DM bin and the shaded region represents 1-99 percentiles (approximate 3$\sigma$). Blue points show observations of Milky Way pulsars using LOFAR done by \citet{Sobey2019} and as queried by \citet{Seta2021} from the ATNF Pulsar Catalogue \citep{Manchester2005}, with the solid blue line showing the median $|\rm{RM}|$ of both sets of observations at each DM, with 30 bins of equal numbers of observations.}
    \label{fig:pulsar_obs}
\end{figure}

In Figure \ref{fig:pulsar_obs} we show the synthetic RM vs. the synthetic DM for m12i, along with MW pulsar data from the Australian Telescope National Facility (ATNF) catalogue \citep{Manchester2005} and from the Low-Frequency Array (LOFAR) \citep[][]{Sobey2019}.  Our simulated RMs and DMs are in good agreement with the observations, with most of the observational measurements towards MW pulsars falling within 1$\sigma$ of the simulation scatter and almost all within 2$\sigma$. This indicates that we find similar magnetic field strengths and geometries averaged over sightlines through the disk in the warm, ionized medium to what is seen in the Milky Way. Between the CR+ and MHD+ runs, we find negligible differences in the inferred properties of the magnetic field in the WIM, consistent with our previous results that CRs do not substantially affect the magnetic properties and observables in the ISM.
  
As many have noted \citep{Simard1980,Rand1989,Han1999,Han2006,Sobey2019}, a naive estimate of $|\langle \rm{B_{\|}} \rangle| \sim 1.2\,|{\rm RM}|/{\rm DM}$ gives $|\langle \rm{B_{\|}} \rangle| \sim 0.1-10\,{\rm \mu G}$ with a median around $\sim 1\,{\rm \mu G}$. Note that $|\langle \rm{B_{\|}} \rangle|$, which is probed by $1.2\,|{\rm RM}|/{\rm DM}$, differs from $\langle |\rm{B_{\|}}|\rangle$ in that the absolute value is taken \textit{after} the line-of-sight averaging. This, in turn, means that $1.2\,|{\rm RM}|/{\rm DM}$ will be sensitive to reversals of the magnetic field along the line of sight, under-predicting the true magnitude of $\langle |\rm{B_{\|}}|\rangle $ and thereby informing us about the coherence of the magnetic field along the line of sight.
  
 In our simulated disks, we see that $1.2\,|{\rm RM}|/{\rm DM}$ under-estimates the actual average values of $\langle |{\bf \rm{B}}| \rangle$ weighted by the thermal electron density by a factor of $\sim 2.5$. Note that $\langle |{\bf B}| \rangle$, which is the linear-weighted average of the magnitude of $\bf \rm{B}$, is a factor of 1.2-2.2 smaller than the "rms"  $\bf B$ often quoted in the literature, for example in \citet{Seta2021}, and is given by $\langle {\bf B^2} \rangle^{1/2}$, where the multiplicative factor depends on the clumping of the magnetic field. Along sightlines drawn through our simulated WIM, this factor appears to be $\sim$ 1.2.
  
 In the next section, we will explore how RMs and DMs trace magnetic fields in the CGM, and how CRs may have an impact on these observables.

\color{black}

 \begin{figure*}
    \centering
    \includegraphics[width=1.0\textwidth]{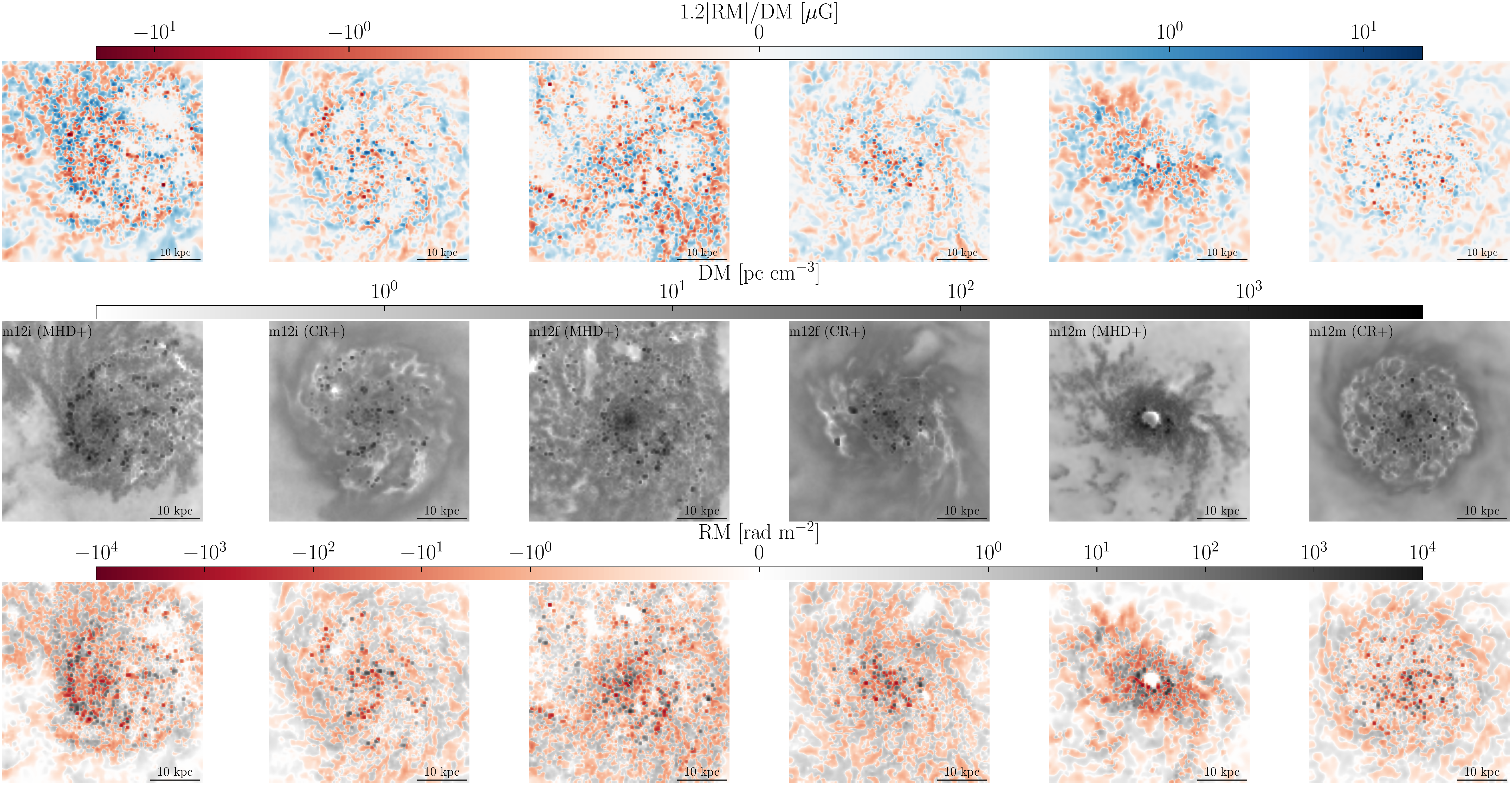}
    \caption{\textit{Visualizations of various line of sight quantities within the central 40 kpc.} \textbf{Row 1:} Line-of-sight RM/DM-inferred magnetic field strength. \textbf{Row 2:} Dispersion Measures. \textbf{Row 3:} Rotation Measures, with m12i (MHD+ and CR+), m12f (MHD+ and CR+), and m12m (MHD+ and CR+) form left to right. Our simulations exhibit small-scale reversals in the sign of RMs similarly to observations, indicative of small-scale field reversals due to explicit treatment of stellar feedback and partially resolved ISM phase structure.}
    \label{fig:20kpcviz}
\end{figure*}

\begin{figure*}
    \centering
    \includegraphics[width=1.0\textwidth]{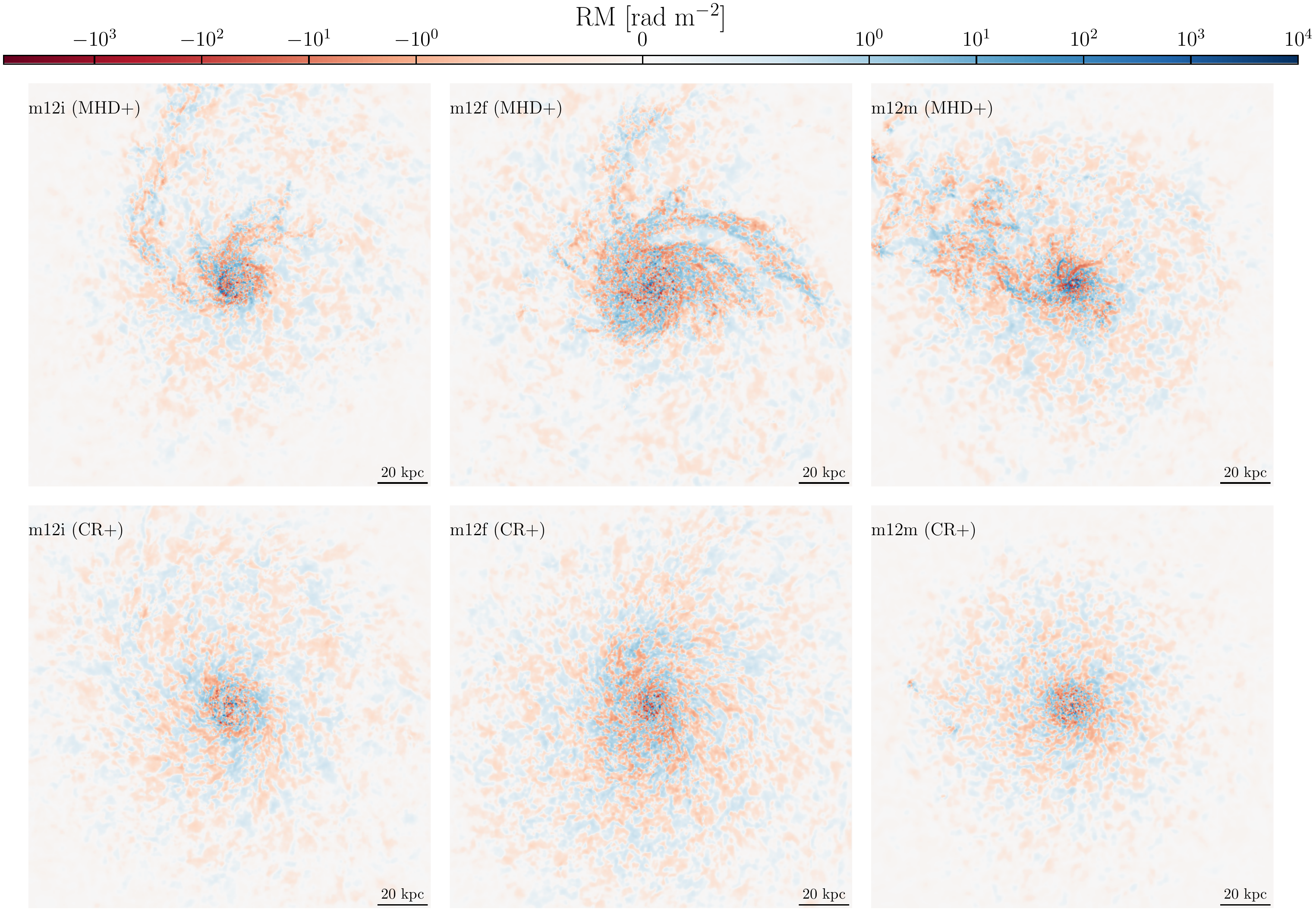}
    \caption{\textit{Face-on visualizations of synthetic rotation measures out to a projected radius of 200 kpc as seen by an external observer towards an ideal background source.} \textbf{Row 1:} MHD+ simulations. \textbf{Row 2:} CR+ simulations, with m12i, m12f, m12m from left to right. Large scale gas inflows and spiral structure can be seen, as well as sign reversals in RM on both large and small scales. In m12m MHD+, an in-falling satellite galaxy and tidal tail can be observed.}
    \label{fig:RM200}
\end{figure*}

\subsubsection{RMs and DMs in the ISM and CGM to distant observers}

Recently, there has been a surge of theoretical and observational interest in the properties (including magnetic fields) of the CGM, so we extend our comparison to galactocentric radii $\sim 10-200\,$kpc. Note that we restrict to RMs and DMs here (as opposed to Zeeman splitting), since the CGM gas is predominantly ionized and there do not exist Zeeman splitting data for the CGM.

In Fig. \ref{fig:20kpcviz} we visualize the RMs, DMs, and $\rm B_{\|}$ seen in a face-on projection of the galaxy zoomed in on the central 40 kpc, while in Fig.  \ref{fig:RM200} we present visualizations  out to 200 kpc. Inspection reveals sign flips (field reversals) in RM or $\rm B_{\|}$ on both large and small scales, with notable features including the spiral arms and large-scale inflows joining the disk -- these features are consistent with observations, as discussed below.

Constraints on the strength and geometry of ISM magnetic fields in nearby galaxies come from intensity and RM measurements of diffuse radio emission \citep{Beck2015,Han2017}. \citet{Fletcher2011} published the most detailed map of RMs, towards the face-on spiral galaxy M51 by inferring RM through modeling the variation of the synchrotron polarization angle with wavelength at 3-6 cm. Qualitatively, the RM visualization is similar, with sign flips on all observed scales. This is also similar to what is inferred from modeling the spatial distribution of MW pulsar RMs and DMs, and variations in dust polarization and synchrotron to construct galactic magnetic field maps \citep[e.g.][]{Jansson2012,Haverkorn2015,Han2017,Beck2019}. The same appears to be true in the LMC, from visual inspection \citep{Gaensler2005}.

Briefly, \citet{Fletcher2011} explicitly note that they cannot measure the shape of the distribution of RMs in M51 (as their observed RM distribution is noise-dominated), but they can estimate the intrinsic rms value of $|{\rm RM}|$ averaged over the Galactic disk (giving $\langle |{\rm RM}|^{2} \rangle^{1/2}\sim 10\,{\rm rad\,m^{-2}}$, about half the measured rms before accounting for observational errors), which is similar to rms values we obtain in our simulated disks in Fig. \ref{fig:20kpcviz} of around $\sim 20\,{\rm rad\,m^{-2}}$.

\begin{figure*}
    \centering
    \includegraphics[width=0.98\textwidth]{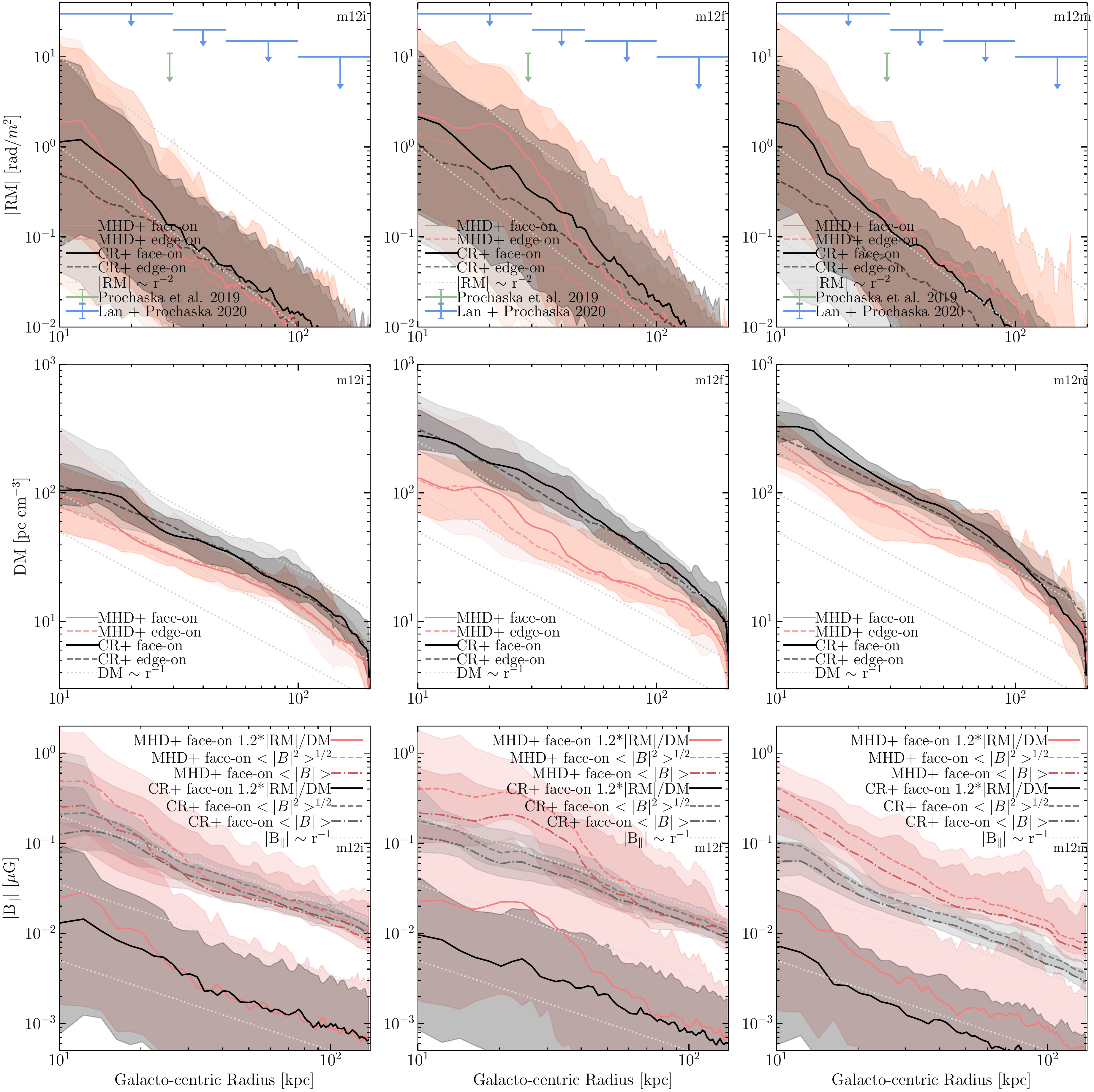}
    \caption{\textit{Rotation measure ($|\rm{RM}|$), dispersion measure (DM), and $|\rm{B_{\rm \|}}|$ profiles as a function of galactocentric radius (R).} \textbf{Row 1:} RM profiles with MHD+ simulations (m12i, m12f, m12m from left to right) shown in coral and CR+ simulations shown in black. Face-on and edge-on profiles are denoted by solid and dashed lines respectively. Each line shows the median at a given radial bin. Shaded regions show 5-95 percentile intervals (approximate 2$\sigma$ error-bars). The green point shows an upper-limit from the RM measured towards an FRB in the halo of a galaxy with M$_*$ $\sim$ 10$^{10.69}$ M$_\odot$ studied by \citet{Prochaska2019a}. The blue lines show 3$\sigma$ upper-limits on RMs in the CGM from a sample of high-redshift radio sources studied by \citet{Lan2020}. Grey dotted lines show representative power-laws as a visual guide. \textbf{Row 2:} Dispersion measure (DM) profiles with the same color and line style conventions as the RM profiles. 
    \textbf{Row 3:} Estimates of $|\rm{B_{\rm{\|}}}|$ as determined from 1.232 RM/DM shown in solid lines, and "true" $<|\rm{B}|>$ and $<|\rm{B}|^{2}>^{1/2}$ shown with dot-dashed and dashed lines repsectively. Our RM and DM profiles are not in tension with the existing observations, which are not yet constraining. The predicted 1.2$\, |\rm{RM}|$/DM profiles are consistent with what would be expected for those of an isothermal sphere. Subtle systematic offsets between MHD+ and CR+ RM profiles exist directly due to the offset in diffuse gas near the disk at low R, and far from the disk in the DM profiles due to CR pressure support. Our $|\rm{B_{\rm{\|}}}|$ profiles indicate that RM/DM significantly under-predicts $<|\rm{B}|>$ averaged over large lines of sight as well as the rms $<|\rm{B}|^{2}>^{1/2}$ by a factor of $\sim$ 15-20.}
    \label{fig:RM_profiles}
\end{figure*}

We quantify the dependence of various quantities of interest as a function of galactocentric radius in the CGM in Figure \ref{fig:RM_profiles}. We present cylindrically averaged radial profiles of RM, DM, and $\langle |{\rm B}|^{2} \rangle^{1/2}$ in the CGM. A few trends are evident. First, we see that while there is considerable detailed spatial structure in Fig. \ref{fig:RM200}, the cylindrically-averaged median profiles (Fig. \ref{fig:RM_profiles}) are quasi-universal radial power-laws in impact parameter $R$, with median ${\rm DM} \propto R^{-1}$ (expected for gas in an isothermal-sphere type profile with three-dimensional $\rho \propto r^{-2}$), $\langle |B_{\|}| \rangle \propto R^{-1}$ and correspondingly $|{\rm RM}| \propto R^{-2}$, on average. The range of slopes for each is roughly $\pm 0.3$. The intrinsic scatter about the trend is large, however, $\sim 1$\,dex in $|RM|$, and subsequently similar in $1.2 \, |RM|/DM$. Interestingly, the trend in $\langle |\rm{B_{\|}}| \rangle$ is a bit shallower than what we might expect for isotropic flux-freezing ($|{\bf B}| \propto \rho^{2/3} \propto R^{-4/3}$), closer to what we might expect for $\langle |\rm{B_{\|}}| \rangle \propto \rho^{1/2}$ in the CGM, similar to the trend seen in the ISM, in Fig. \ref{fig:Bvsdens}.

Second, there are some small but systematic offsets between the MHD+ and CR+ simulations. In the inner CGM approaching the disk ($r\sim 10\,$kpc), $\rm{B_{\|}}$ is a factor of 1.5-2 higher in MHD+, directly related to the offset in Fig. \ref{fig:Bvsdens_wo_obs}. Far from the disk, the DMs are a factor of 1.5-2.5 lower in MHD+: this owes to the lack of CR pressure (able to support a larger "weight" of CGM gas) and inefficient galactic outflows leading to more accretion from the CGM onto the galaxy \citep{Ji2020,Ji2021,Hopkins2021}. In contrast, within the disk, this leads to higher DMs for MHD+. Together, these mean that while typical ISM RMs are larger in the more massive, more dense MHD+ simulations, \textit{the RM profile in the CGM is nearly identical}.   

Third, we see that the inferred $\langle \rm{B_{\|}} \rangle \sim 1.2\,|{\rm RM}|/{\rm DM}$ under-estimates the median $|\rm B_{\|}|$, or $\langle | {\bf B}_{\|}| \rangle$ of gas along the LOS by a factor of $\sim 5-7$. The fields are close to isotropic in this statistical and cylindrically-averaged sense, where we are averaging multiple lines of sight along annuli at a given galactocentric radius. So, this means that $\langle |{\bf B}|\rangle$ $\sim$ $2 \langle | {\bf B}_{\|}| \rangle$, which means on average $|{\rm RM}|/{\rm DM} \sim 0.1\,\langle |{\bf B}|\rangle$ where the averaged terms are weighted by electron density in the same manner as RM. We have confirmed that this owes primarily to cancellation due to random field components along the LOS; crudely, this systematic offset is equivalent to the statement that the coherence length of the magnetic field is $\sim 10\%$ the size of the system contributing to RM (i.e. $\sim 20\,$kpc -- comparable to the halo scale length -- in the outer CGM). 

While there is no observational measurement of RM from the CGM around any $\sim L_{\ast}$ galaxy, a number of studies measuring RMs from background fast radio bursts (FRBs) and other bright radio sources have placed upper limits which can be compared to the predictions of our simulations. At the radii shown in Figure \ref{fig:RM_profiles}, we plot the most stringent upper limits to date, specifically the upper limit towards an FRB found in \citet{Prochaska2019a} and the $3\sigma$ upper limits quoted as a function of impact parameter in the FRB study of \citep{Lan2020}.\footnote{Note that while the FRBs used in \citet{Prochaska2019a,Lan2020} have measured RMs, these RMs are almost certainly strongly dominated by the contribution from the FRB host galaxy, with an unknown additional contribution from the FRB source and IGM, so the authors can place only a statistical upper limit on the contribution to the measured RM from the CGM of the foreground $\sim L_{\ast}$ galaxy.} At even larger impact parameters $\gtrsim 500\,$kpc (not shown), \citet{Ravi2016} place a $2\sigma$ upper limit on $\langle {\bf B_{\rm \|}}\rangle_{\rm IGM}$ of $2.1\times 10^{-2}\,{\rm \mu G}$, and \citet{O'Sullivan2020} measure a $2\sigma$ upper limit $|{\rm RM}| < 1.9\,{\rm rad\,m^{-2}}$ and constrain the IGM magnetic field strength to be $\langle |{\bf B}|\rangle_{\rm IGM} < 4\times10^{-3}\,{\rm \mu G}$, both from FRB detections. We see that the simulations are easily consistent with all of these limits, but the data are not yet particularly constraining. Still, with much larger FRB samples expected in the near future from DSA-110, CHIMES, and CHORD it should be possible to improve these upper limits by an order of magnitude or more, potentially reaching detection thresholds at least in the inner CGM \citep{Vanderlinde2018,Amiri2018,Kocz2019,Connor2021}. Further observational data are needed to delineate between differing physical and numerical schemes which predict order of magnitude differences in halo field strengths from our results, discussed below \citep{Pakmor2020a}.

\section{Discussion and Conclusions}\label{sec:discussion}
The results in this work have presented probes of the magnetic field as traced in different gas phases; cold, dense and neutral ISM gas (T $\sim$ 100 K, $\rm n_H$ $\gtrsim$ 10 cm$^{-3}$), warm ionized ISM gas ($\rm n_H$ $\sim$ 0.1 cm$^{-3}$, T $\gtrsim$ 6000 K), and hot, diffuse CGM gas (T $\gtrsim$ 10$^{6}$ K, $\rm n_H$ $\sim$ 0.001 cm$^{-3}$). Due to having resolved phase structure of the ISM in these simulations, we are able to meaningfully compare these tracers to their observational counterparts without use of additional modeling or assumptions, and are thus able to both (a) test whether observational assumptions apply in these simulations and (b) comment on how two physical models (MHD+, CR+) compare in their predictions of the observables.

We investigated how the magnetic field properties vary in two types of simulations of the same galaxies: MHD+ and CR+. We find general agreement in the scaling of the overall relation of $\rm |\textbf{B}|$ vs. number density in all of the simulations, but with a modest offset between the MHD+ and CR+ runs of a factor of $\sim 2-10$ primarily at very diffuse ISM densities (n $<$ 0.01 cm$^{-3}$) which increases towards more diffuse densities and is more systematic for m12m, with an offset of a factor of $\sim$ 2 at (n $>$ 1 cm$^{-3}$). This difference at very diffuse ISM densities, as mentioned above, arises primarily due to differences in the "inner CGM." 

The "inner CGM" is precisely where previous studies \citep{Booth2013,Simpson2016, Girichidis2018, Buck2020, Ji2020,Ji2021,Hopkins2021,Chan2021} have shown CRs can play a dramatic role influencing the dynamics of galactic fountains and outflows. Specifically, in simulations without cosmic rays, MW-mass galaxies at low redshifts have their outflows "trapped" by CGM pressure, creating high-velocity fountains with rapid recycling \citep{Muratov2015,AnglesAlcazar2017,Gurvich2020,Stern2021,Hafen2022}, while in the CR+ simulations, the added CR pressure gradient both maintains gas acceleration and reduces the pressure barrier to outflows, allowing disk outflows to smoothly escape to the outer halo (references above). We directly confirm that, as a result, the velocity dispersion of the low-density inner-CGM gas which drives the offset of $|{\bf B}|$ versus $n_{\rm gas}$ is lower in the CR+ simulations by a factor of $\sim 2-3$. This, in turn, means that magnetic fields experience significantly less amplification by "turbulent" or fountain flows in this range of conditions due to recycling of outflows \citep{Martin-Alvarez2018}, and greater adiabatic attenuation, since the escaping outflow will reduce the magnetic field strength of any advected fields from the disk via flux-freezing as the outflowing gas expands. This can explain some of the offset in $\rm |\textbf{B}|$ vs. number density between MHD+ and CR+ galaxies, but not the whole difference.

From our CR+ simulations, which self-consistently evolve the GeV CRs which dominate the CR energy budget, we are able to make predictions for the relative strengths of magnetic, thermal, and CR pressures in different gas phases. We find that in the disk, equipartition assumptions between CRs and magnetic fields  \citep[as is often assumed for estimating magnetic field strengths from observations of diffuse radio emission][]{Chyzy2011,Beck2015} may hold on large scales, and at typical ISM densities (L $>$ 1 kpc, $n_{\rm H}$ $\sim$ 1-10 cm$^{-3}$); however in the CGM, this assumption breaks down. Similarly, assumptions of equipartition between thermal and magnetic pressures in the CGM appear to be invalid given the predictions of our model; however, details of this depend on how CRs are treated and their interplay with metal-enriched, actively-cooling gas \citep{Prochaska2019a,Hopkins2021c}. That being said, in both the halo and the ISM, we generally see a plasma $\beta \ll$ 1 in cold/neutral dense gas (reaching values as low as 5$\times$10$^{-5}$), and $\beta \gg$ 1 in low density, mostly-ionized gas (reaching values as high as 7$\times$10$^{8}$), consistent with most previous theoretical studies \citep{Su2018, Ji2020, Butsky2020}.

We compare ISM magnetic field values inferred from Zeeman measurements within the Galaxy for the cold, neutral medium, and find broad agreement (Figure \ref{fig:BvsNH}). In detail, there are some differences between runs, primarily that the offset in magnetic field strength at lower gas densities results in CR+ simulations predicting slightly lower values of B$_{\|}$ relative to the MHD+ simulations at lower column densities, which probe more of the diffuse gas. This results in slightly poorer agreement at lower column densities (log$_{10}$(N$_{\rm{H}}$) < 21 cm$^{-2}$) between the existing observations of \citet{Crutcher2011} and our simulations. While the CR+ simulations exhibit slightly lower values of B$_{\|}$ relative to the MHD+ simulations, we note that the overall qualitative agreement with the observations is not notably different between the two physical models. 

The physical relations of B$_{\|}$ vs. density (Figure \ref{fig:Bvsdens}) for both the MHD+ and CR+ simulations agree considerably well with the Solar circle diffuse \ion{H}{I} observations of \citet{Crutcher2010}. The observations of interest span higher densities (n $>$ 300 cm$^{-3}$), where there is little difference in the magnetic field strength of the MHD+ and CR+ simulations. 

As discussed in Section 3.3.2, the poor correlation between n$_{\rm HI,T_{CNM}}$ and n$_{\rm HI,M_{CNM}}$ flattens the relation between $|B_{\rm \|}|$ and n. This may explain some aspects of the \citet{Crutcher2010} analysis, which compiles Zeeman observations from three types of clouds (diffuse \ion{H}{I}, IR dark, and dense molecular) of varying densities using three different tracers (\ion{H}{I}, OH, CN). It is important to note that {\em within any one sample}: i.e. within any one "type of cloud" {\rm or} within any one tracer, there is no statistically significant trend of $|\rm{B_{\|}}|$ with $n_{\rm HI,\,T_{\rm CNM}}$ in the \citet{Crutcher2010} data. This leads, for example, to their conclusion that the Zeeman-inferred $|\rm{B_{\|}}|$ is approximately constant in diffuse gas with $n < 300\,{\rm cm^{-3}}$, which are exclusively sampled by diffuse \ion{H}{I} -- it is only when the different datasets, each of which samples different median density ranges (e.g. diffuse \ion{H}{I} at $10^{1}-10^{2.5}\,{\rm cm^{-3}}$, dark clouds for $10^{3.5}-10^{4.5}\,{\rm cm^{-3}}$, dense molecular clumps at $10^{5.5}-10^{6.5}\,{\rm cm^{-3}}$) are combined that the underlying trend (quite similar to what we predict here) can be observed. Here, we show that detecting the predicted trend {\em within} the dynamic range of densities probed by \ion{H}{I} Zeeman data alone (a factor of $\sim 10-30$ in gas density) requires a three-dimensional gas density estimator which is accurate to much better than this dynamic range (i.e. to within a factor of $\sim 2$ or so).

In visually examining the synthetic rotation measures for our simulated galaxies, we find variations of the sign of RM on small scales as well as large scales, with quantitative values in agreement upper limits placed in the halos of L$_*$ galaxies by \citet{Prochaska2019a,Lan2020}. Zooming in on the disk, we find RM sign reversals on small scales in a manner qualitatively similar to the RM maps of M51 produced by \citet{Fletcher2011}, with little to no correlation of the RMs with galactic structure, differing from work done by \citet{Pakmor2018}, who performed a similar study of the RM by analyzing a MW-analog from the Auriga simulations. Our synthetic RMs show a less ordered, more turbulent magnetic field in better agreement with the observations, which is primarily indicative of the effect of explicit stellar feedback and resolution of the ISM phase structure, in contrast to the sub-grid ''effective equation-of-state'' model used for the ISM in that work.

For the warm, ionized phase of the ISM of our simulated galaxies, we find good agreement with RMs and DMs towards Milky Way pulsars. While the mean RM/DM-inferred magnetic field strength ubiquitously under-predicts the mean "true" rms $|{\bf B}|$ in the disk, at the column densities of interest, it traces B$_{\rm{\|}}$ averaged along sightlines through the disk.

While we primarily focus the discussion of the RMs, DMs, and subsequent inferred estimates of B$_{\rm{\|}}$ on the halo in Figure \ref{fig:RM_profiles}, we find good qualitative agreement between the intrinsic dispersion of the RM as inferred from synchrotron polarization at 3 and 6 cm by \citet{Fletcher2011}, with the caveat that the methodology of our synthetic RMs does not aim to faithfully reproduce that of \citet{Fletcher2011} and contains information averaged over large ($\sim$ 280 kpc deep) sightlines rather than solely arising from the disk. In the region that is primarily of interest for these synthetic background point source RMs, i.e., the halo, the MHD+ and CR+ profiles converge, independently of resolution. 

Our CGM-focused results suggest that use of RMs and DMs towards point sources as measures of B$_{\rm{\|}}$ underestimate the "true" magnetic field strength in the halo at a given galactocentric radius by about a dex. This is consistent with work done by \citet{Seta2021}, who found that in the presence of driven subsonic, transonic, and supersonic turbulence, the standard deviation of the average parallel component of the magnetic field is about an order of magnitude less than the true rms magnetic field strength of the box. Here, in Figure \ref{fig:RM_profiles}, the RM/DM estimate traces effectively the ionized-mass-weighted parallel component of the magnetic field, and $<|\rm{B}|^{2}>^{1/2}$ averaged over the
extensive (~240 kpc) lines of sight probes the root mean square magnetic field strength. This result may be of particular importance for observers looking to characterize the magnetic field strengths in the halos of galaxies towards background FRBs, implying that RM/DM estimates may be a factor of $\sim$ 10 lower than the true halo magnetic field strength. From our estimate of how well the RM/DM-inferred B$_{\rm \|}$ traces the "true" rms $\langle |\bf \rm{B}|^{2}\rangle ^{1/2}$ we infer that the coherence length is $\sim$50$\%$ the characteristic length scale in the ISM (on order of the disk scale height), and $\sim$10-20$\%$ of the characteristic length scale in the CGM, on order the halo scale length. This result on the coherence length is similar to predictions or the large-scale coherence length towards MW pulsars from random walk models described by \citet{Seta2021}.

Notably, the predictions of RMs and $\langle|\bf \rm{B}|^{2}\rangle^{1/2}$ of the Auriga cosmological zoom-in simulations presented in \citet{Pakmor2020a} are a factor of $\sim$ 10 higher than those computed from the FIRE simulations. The Auriga simulations use an effective equation-of-state model of the ISM \citet{Springel2003} in contrast to the explicit treatment of feedback used in our simulations. The typical inter-cell spacing in the CGM of the halos of FIRE and Auriga are similar ($\sim$ kpc), and both are notably turbulent \citep[see][]{Ji2020}, however it remains unclear what dominates magnetic amplification in the halo, and whether resolving the inertial scale of turbulence in the CGM will be key to generating accurate halo magnetic fields. We note also that varying the $\sigma_{p}$, $\sigma_{h}$, and $\alpha_{\psi}$ divergence cleaning and slope-limiter terms \citep[see]{Hopkins&Raives2016} by factors of $\sim$5 does not affect the magnetic field properties of these simulated galaxies, indicating that the magnetic field strength is set by physical processes rather than numerics. We have shown that in our simulations with resolved ISM phase structure, the magnetic field strengths very reasonably agree with existing constraints from the Milky Way and M51, however the halo remains uncertain. Present observational constraints are unable to distinguish between these different physical models and numerical treatments, and future observations \citep[e.g.][]{Kocz2019} will be key for understanding CGM magnetic fields.

\section{Summary and Future Work}\label{sec:conclusions}
In this paper, we have analyzed the magnetic fields in the multiphase ISM and CGM in a set of high-resolution, cosmological simulations from the FIRE-2 project, run with two different physical models (MHD+, CR+). We analyze the differences between the magnetic fields in simulations run with each model, and compare forward-modeled observables tracing magnetic fields in the ISM and CGM to existing observational constraints.

In summary, our conclusions are as follows:
\begin{itemize}
    \item Inclusion of CRs in simulations of galaxy formation does not directly affect the magnetic field strengths in the ISM at a given gas density, though there are modest differences in the lowest density gas (n $<$ 0.01 cm$^{-3}$) due to dynamical effects of CRs in the disk-halo interface (the inner most region of the CGM). The main effect of CRs is indirect, causing the average $|B|$ in those galaxies' less massive, less dense disks to be lower than their MHD counterparts, moving along the same scaling relations between $|B|$ and $\rm n_H$.
    
    \item Our predicted relation for $|\bf \rm{B}|$ vs. three dimensional gas density $\rm n_H$ is consistent with that of flux freezing in spherically symmetric and non-spherical geometry ($|\bf \rm{B}| \sim \rm n_H^{2/3}$ or $|\bf \rm{B}| \sim \rm n_H^{1/2}$).
    
    \item Equipartition between magnetic, thermal, and cosmic ray pressures is achieved in our simulated galactic disks at typical ISM densities (n $\sim$ 1-10 cm$^{-3}$). Equipartition between magnetic and cosmic ray energy densities primarily holds on large ($>$ 1 kpc) scales, cospatial with the spiral structure of our galactic disks, and fails in the interarm regions and on small scales in the presence of dense molecular gas. 
    \item In the halos of our simulated galaxies, equipartition between magnetic pressures and cosmic ray pressures does not hold, and neither does equipartition between thermal and magnetic pressures ($\beta >> 1$).
    
    \item Our simulated magnetic field strengths in the cold, neutral ISM agree well with existing Zeeman observations in the Milky Way, and indicate observational estimates of the three dimensional gas density in diffuse HI from spin temperatures may be noisy due to thermal pressure equilibrium being a poor assumption in this phase of the ISM. This can act to obfuscate existing correlations between $|\bf \rm{B_{\rm |}}|$ and $\rm n_H$ for $\rm n_H <$ 300 cm$^{-3}$.
    
    \item The magnetic field strengths in the warm, ionized ISM of our simulated galaxies are in agreement with observations of Milky Way pulsars and M51 as inferred through rotation measures and dispersion measures.
    
    \item Our synthetic rotation measures as a function of impact parameter are in agreement with existing constraints from FRBs probing the halos of L$_{\ast}$ galaxies, however there is a large parameter space occupied in this plane by various physical and numerical models. Future surveys which will localize 100s of FRBs per year will be crucial to constraining current model predictions.

    \item In the CGM, our CR+ simulations exhibit slightly enhanced DMs relative to the MHD+ simulations in the inner CGM (R $\sim$ 50-100 kpc) due to cosmic ray pressure support.
    
    \item Our simulations' comparison of  1.2 $\, |\rm RM|$/DM ($\langle \bf \rm{B_{\rm \|}} \rangle$) to the "rms" field $\langle |\bf \rm{B^{\rm 2}}|^{1/2} \rangle$ indicates that observational estimates of magnetic fields in the halos of L$_{\ast}$ galaxies from FRBs may under-predict the true rms field strength by a factor of 15-20, in qualitative agreement with previous works.

\end{itemize}

Our study provides motivation to more closely study the magnetic fields in these simulations (e.g., turbulent magnetic field amplification and its connection to feedback models), and to generate detailed mock observations using radiative transfer codes like \texttt{POLARIS} \citep{Reissl2016}. This will enable detailed comparison with more indirect magnetic field tracers, such as synchrotron intensities, magnetic field morphologies inferred from polarized dust and synchrotron emission \citep{Borlaff2021a}, and resolved Zeeman spectra. 

We will also explore predictions for different galaxy types (e.g. dwarfs or starbursts), and redshift evolution, which may help shed light on magnetic field amplification mechanisms. It is also important to continue to explore different physics: as we noted above, our CR+ models adopt a highly simplified empirical (constant-diffusivity) assumption for CR transport, and previous work \citep{Hopkins2021b} has shown that other observationally-allowed models with variable diffusivity can produce different result (so even if CRs are present, reality may closely resemble our MHD+ simulations, for example). 

Additionally, our simulations neglect the effects of Active Galactic Nuclei (AGN), which may be an important component of the feedback budget for L$_*$ galaxies \citep[e.g.,][and references therein]{Wellons2022,Feldmann2022}. Further understanding of galactic magnetic fields and their tracers from a theoretical perspective may lead to insight on questions related to survival and infall of cool CGM gas, star formation efficiency in dense gas, and the ability of cosmic rays to influence galactic properties via transport along magnetic field lines.

\section{Acknowledgements}

We wish to recognize and acknowledge the past and present Gabrielino-Tongva people and their Indigenous lands upon which this research was conducted. Additionally, we thank the staff at our institutes, without whose endless efforts this work would not be possible during the ongoing pandemic. Support for SP and PFH was provided by NSF Research Grants 1911233, 20009234, 2108318, NSF CAREER grant 1455342, NASA grants 80NSSC18K0562, HST-AR-15800. GVP acknowledges support by NASA through the NASA Hubble Fellowship grant  \#HST-HF2-51444.001-A  awarded  by  the  Space Telescope Science  Institute,  which  is  operated  by  the Association of Universities for Research in Astronomy, Incorporated, under NASA contract NAS5-26555. Numerical calculations were run on the Caltech compute cluster "Wheeler," allocations AST21010 and AST20016 supported by the NSF and TACC, and NASA HEC SMD-16-7592. The Flatiron Institute is supported by the Simons Foundation.

\section*{Data Availability}

The data supporting the plots within this article are available on reasonable request to the corresponding author. A public version of the GIZMO code is available at \url{http://www.tapir.caltech.edu/~phopkins/Site/GIZMO.html}. FIRE-2 simulations are publicly available \citep{Wetzel2022} at \url{http://flathub.flatironinstitute.org/fire}, though simulations including the physics of MHD and cosmic rays like those analyzed in this study are not yet publicly available. Additional data, including initial conditions and derived data products, are available at \url{https://fire.northwestern.edu/data/}.



\bibliographystyle{mnras}
\bibliography{B_fields} 

\bsp	
\label{lastpage}
\end{document}